\documentclass[letter,scriptaddress,twocolumn, showkeys]{revtex4}
\usepackage{ amssymb }
	\usepackage{amsmath}%,amssymb} 
	\usepackage{makeidx}
	\usepackage{amsfonts}
	\usepackage[ansinew]{inputenc}
	\usepackage[usenames,dvipsnames]{pstricks}
	\usepackage{subfigure}
	\usepackage{epsfig}
	\usepackage{pst-grad} % For gradients
	\usepackage{pst-plot} % For axes
		\usepackage{amsthm}
\usepackage{ amssymb }
 	\usepackage{makeidx}
	\usepackage{amsfonts}
 	\usepackage{listings}
 	\usepackage{mathtools}
	\usepackage{float}
	
	\usepackage[colorlinks,hyperindex]{hyperref}
	\hypersetup
	{
		colorlinks,%
		citecolor=blue,%
		linkcolor=blue,%
		urlcolor=blue,%
	}
	\usepackage{tikz}

\usepackage{amsmath}
\usepackage{amssymb}
\usepackage{graphicx}
\usepackage{xcolor}
\usepackage{float,graphicx}
\usepackage{placeins}
\usepackage{color, colortbl}
\usepackage[colorinlistoftodos]{todonotes}
\theoremstyle{definition}
	\newtheorem{definition}{Definition}
	\newtheorem{example}[definition]{Example}
	
	\theoremstyle{plain}

\makeindex

\begin{document}

\title{Cell fusion through slime mold network dynamics}

\author{ Sheryl Hsu$^{a}$ and   Laura P. Schaposnik$^{\star, b}$ }
% \affiliation[label1]{}
  \affiliation{($\star$) Corresponding author: schapos@uic.edu}

\begin{abstract}
\textit{Physarum Polycephalum} is a unicellular slime mold that has been intensely studied due to its ability to solve mazes, find shortest paths, generate Steiner trees, share knowledge, remember past events, and the implied applications to unconventional computing. The CELL model is a unicellular automaton introduced in \cite{plu8} that models \textit{Physarum}'s amoeboid motion, tentacle formation, maze solving, and network creation. In the present paper, we extend the CELL model by spawning multiple CELLs, allowing us to understand the interactions between multiple cells, and in particular, their mobility, merge speed, and cytoplasm mixing. We conclude the paper with some notes about applications of our work to modeling the rise of present day civilization from the early nomadic humans and the spread of trends and information around the world. Our study of the interactions of this unicellular organism should further the understanding of how \textit{Physarum Polycephalum} communicates and shares information.
 \end{abstract}

\maketitle
 
\section{Introduction}
 
   \textit{Physarum Polycephalum} is a unicellular slime mold. At one point in its life cycle, it forms a plasmodium with many tubes. Experiments have shown that \textit{Physarum Polycephalum} in this plasmodium state is able to solve mazes, find the shortest path, build high-quality networks between multiple points, adapt and respond to stimulus, and also remember past events \cite{plu14, plu15, plu16}. Recently, several mathematical models  inspired by \textit{Physarum} have been developed to approach problems such as the shortest path or Steiner tree problems \cite{plu8,plu12,plu6,april2}.
   \par
      The behavior of \textit{Physarum Polycephalum} can be studied, in particular,   using cellular automaton as done in \cite{plu8} through what the authors called {\it the CELL model}. In this model, a \textit{Physarum} cell is modeled on a grid with cytoplasm and cytoskeleton. At each iteration, the cytoplasm and cytoskeleton is slightly reshaped with the introduction of an outside bubble. As shown in \cite{plu8}, this model can be used to simulate ameboid motion with tentacle formation, maze solving, shortest path finding, and network creation. 
%      More recently,  \textit{Physarum Polycephalum} has contributed a lot to unconventional computing. In particular, it has been used in \textit{Physarum} learning computer chips in which \textit{Physarum} is grown on an electrode grid and stimulated. Similarly, there has been work done in designing logic gates and chips using \textit{Physarum} \cite{intro1, intro2}. 

Inspired by  \cite{plu8},  we take a novel approach which allows us to discern several interesting patterns. When considering cells, one expects to observe certain behaviors. In particular:
\begin{itemize}
     \item[(a)]   The time it takes for two cells to merge should increase directly with the distance between them.    
     \item[(b)]   It should be  fastest and most likely for cells to merge with the cell closest to them.
    \item[(c)] The smaller a cell is, the more mobile it should be. 
    \item[(d)]   The smaller cells are, the better they should fuse in terms of cytoplasm mixing. 
    \end{itemize}
     
We shall begin this paper by introducing in Section \ref{back} some of the different models used in the literature to study \textit{Physarum Polycephalum}, paying particular attention to the {\it flow-conductivity model} \cite{plu6}, the {\it cellular model} \cite{plu8},  the {\it multi-agent model} \cite{plu12}, and the {\it shuttle-streaming model} \cite{april2}. 

After introducing these models, we dedicate Section \ref{content} to the main findings of our work. 
By considering cells at different distances apart and studying   the number of iterations it took for them to merge, we could see that: 
\begin{itemize}
\vspace{-0.05 in}

\item  The relationship between distance and iterations is linear,  as shown in Figure \ref{fig:linfit_mergedis}.
\end{itemize}
%\vspace{-0.05 in}
By considering three cells and adjusting the distance between them, we measured which cells merge first and the number of iterations it took them to do so, allowing us to see that:
\begin{itemize}
\vspace{-0.05 in}

\item  The two closest cells are most likely to merge together, followed by the next two closest cells, followed by the two furthest cells;
\vspace{-0.05 in}

\item The two closest cells take the least number of iterations to merge, followed by the next two closest cells, followed by the two furthest cells.
 \end{itemize}
 We also considered cells of different sizes and measure mobility using the average distance from spawn point, maximum distance from spawn point, and speed, leading to the following results:
\begin{itemize}
\vspace{-0.05 in}

\item The relationship between average distance and size is exponentially decreasing;
\vspace{-0.05 in}

\item The relationship between maximum distance and size is exponentially decreasing;
\vspace{-0.05 in}

\item The relationship between speed and size is also exponentially decreasing.
\end{itemize}
Finally, by considering the fusion of two cells of varying sizes, we   measured cytoplasm mixing using Lacy's mixing index \cite{plu13}, leading to the following findings:
\begin{itemize}
\vspace{-0.05 in}

\item There is a decreasing exponential relationship between the mean mixing index and cell area;
\vspace{-0.05 in}

\item There is a decreasing logistical relationship between max mixing index and cell size.
\end{itemize}
We conclude this paper  expanding on the analysis and applications of the above findings in Section \ref{last}. In particular, we explore the applications of our work to modeling the rise of present day civilization from the early nomadic humans and the spread of trends and information around the world.
\newpage

\section{Modelling {Physarum Polycephalum}}\label{back}
  We shall dedicate this section to reviewing four models inspired by \textit{Physarum Polycephalum},  and which we shall consider in the present paper when building a new model to study cell fusion between multiple cells. We shall begin by reviewing a flow-conductivity model in Section \ref{cond}, a cellular model in Section \ref{celu},  and a  multi-agent model in Section \ref{multi}.  Finally, we shall conclude the section by introducing a shuttle-streaming model in Section \ref{shut}. A good overview of current literature on {\it Physarum Polycephalum} can be found at \cite{plu3}.

  \subsection{Flow-conductivity model}\label{cond}
 
Inspired by  the idea that within \textit{Physarum Polycephalum} tubes of the plasmodium grow thicker as more protoplasm flows through that tube, the authors of \cite{plu6} introduced a flow-conductivity model which we shall review below. In this setting, two further observations about \text{Physarum}'s behavior come into play: 
\begin{itemize}
\item when two tubes connect to the same food source, the longer one usually disappears;
\item open-ended tubes are likely to disappear.
\end{itemize}
 
    In what follows, we shall describe this model following the ideas of \cite{plu6}. Consider nodes $N_1, N_2 \dots$, and let $N_1$ be the source node (start) and $N_2$ be the sink node (end). The tube/edge connecting nodes $N_i$ and $N_j$, is denoted as $M_{ij}$. Let $Q_{ij}$ be the flux through $M_{ij}$, let $p_i$ be the pressure at node $i$, let $L_{ij}$ be the length of $M_{ij}$, and let $D_{ij}$ be the conductivity of $M_{ij}$.
  Then, we have 
  \begin{equation}
  Q_{ij} = \frac{D_{ij}}{L_{ij}}(p_i - p_j).\label{qi}
  \end{equation}
  By Kirchhoff's law,
  \begin{eqnarray}\label{2}
  \sum_{i} Q_{ij} = 0 \quad (j \neq 1, 2).
  \end{eqnarray}
  Since $N_1$ and $N_2$ are the start and end node respectively,
  \begin{eqnarray}
  \sum_{i} Q_{i1}+I_0 &=& 0;\\
   \sum_{i}Q_{i2}-I_0&=&0.\label{4}
  \end{eqnarray}
 
 The adaptation of the tube thickness  can be modeled using the following equation:
  \begin{equation}
  \frac{d}{dt}D_{ij} = f(|Q_{ij}|)-rD_{ij},
  \end{equation}
  where $f(Q)$ must be an increasing function with $f(0) = 0$, and thus for the purposes of this model one may consider $f(Q) = \alpha|Q|$.
Setting $p_2 = 0$, one can compute all the $p_i's$ using equations \eqref{2}-\eqref{4} and therefore all $Q_{ij}$.
 The flow-conductivity model has been used to find Steiner trees\cite{6684158}, solve the traveling salesman problem \cite{10.1007/978-3-642-38703-6_35}, and design fault-tolerant networks \cite{FaultTolerant}. It has been applied to creating a shortest path navigation system across the US interstate highway \cite{plu6}, designing railroad networks similar to the Tokyo railroad system \cite{Tero439}, designing transportation networks with changing traffic distributions  \cite{10.1371/journal.pone.0089231}, identifying focal nodes for disease spread in epidemiological networks \cite{disease}, and solving supply chain network design problems \cite{doi:10.1080/00207543.2016.1203075}.
  %This model is used in \cite{plu6} to create a shortest path navigation system across the US interstate highway. By setting one terminal to the sink node and the others to source nodes, this model was also applied to Steiner trees. Moreover, fast algorithms inspired by \textit{Physarum Polycephalum} have been used for the node weighted Steiner tree problem with multiple terminals, and it has also been used to design transportation networks, focal nodes in disease spread in epidemiological networks, spectroscopy analysis, and drug repositioning \cite{back2}.    Finally,  another application  of these types of models is finding node placement in wireless sensor networks to minimize cost \cite{back1}.

 \vspace{0.1 in}
 \hrule
 
 \vspace{-0.05 in}
  
\begin{example}In order to illustrate how the {\it flow-conductivity model} can be used, consider the  graph in Figure \ref{col2}, where 1 is the source node and 6 is the sink node. 
\vspace{-0.3 in}
             \begin{figure}[H]
  \centering \includegraphics[scale = 0.15]{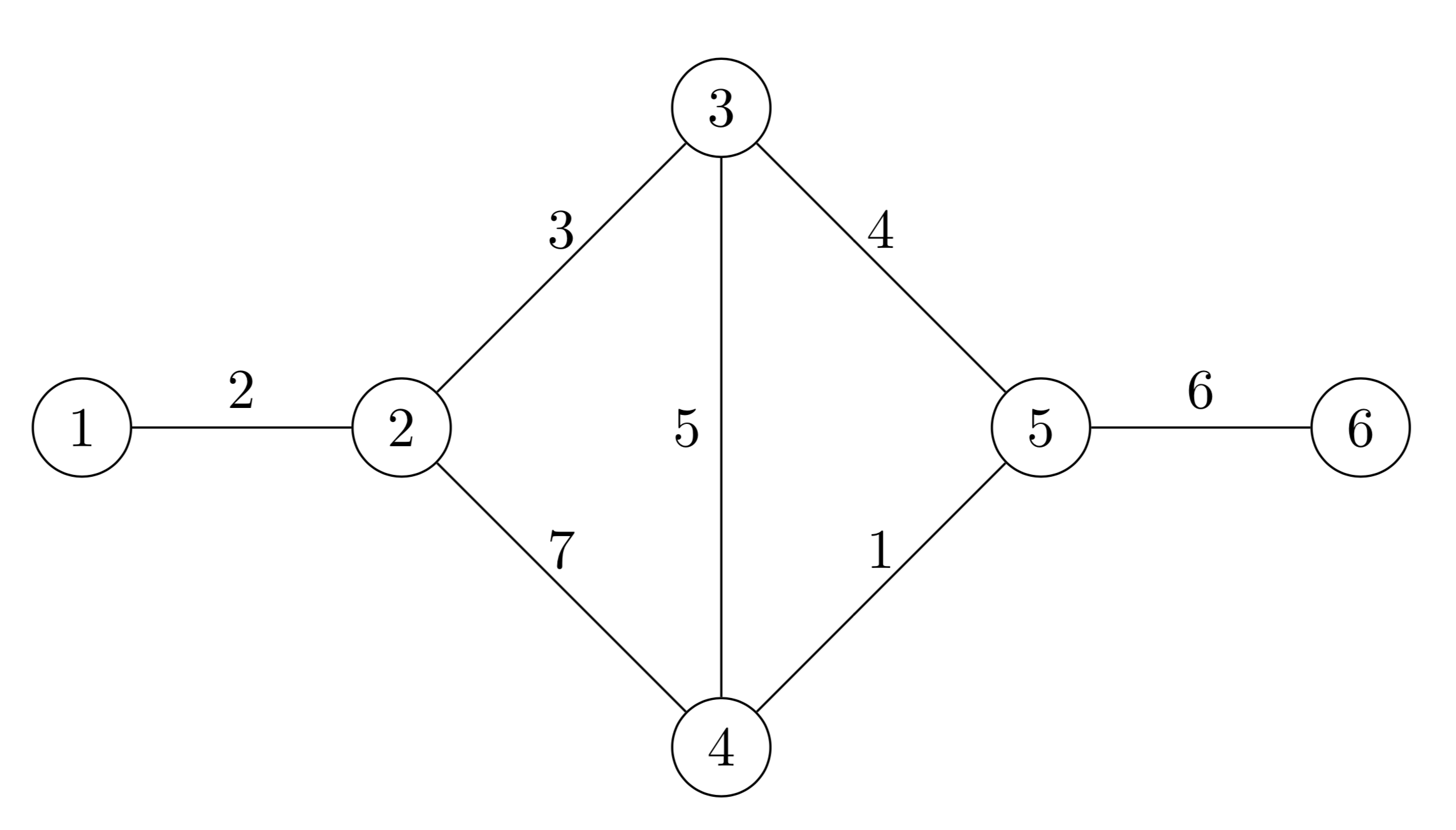}
  \caption{Node 1 is the source node and  node 6 is the sink node. The number on each edge represents the length of that edge.}
      \label{col2}
\end{figure}

Starting  with  a flux of $I_0 = 10$ into node 1, and  assuming  all tubes start off with the same conductivity  $D_{ij} = 2$,  one has the following system of equations:
\begin{eqnarray}
	\frac{2}{2}(p_2 - p_1) + 10 &=& 0
\nonumber\\
	\frac{2}{2}(p_1 - p_2)+\frac{2}{3}(p_3-p_2)+\frac{2}{7}(p_4 - p_2) &=& 0
\nonumber \\
	\frac{2}{3}(p_2 - p_3)+\frac{2}{5}(p_4 - p_3)+\frac{2}{4}(p_5 - p_3) &=& 0\nonumber\\
	\frac{2}{7}(p_2 - p_4)+\frac{2}{5}(p_3 - p_4) + \frac{2}{1}(p_5 - p_4) &=& 0
\nonumber\\
	\frac{2}{4}(p_3 - p_5) + \frac{2}{1}(p_4 - p_5) + \frac{2}{6}(p_6 - p_5) &=& 0\nonumber\\
	\frac{2}{6}(p_5 - p_6) - 10 &=& 0\nonumber.
\end{eqnarray}

Setting $p_6 = 0$ and solving this system, leads to the values  $p_1 = 57;$ $p_2 = 47;$ $p_3 = 38;$ $p_4 = 33;$ and  $p_5 = 30.$ Then, from  Eq. \eqref{qi}
%$$Q_{ij} = \frac{D_{ij}}{L_{ij}}(p_i - p_j).$$
one has  $Q_{12} = 10;$ $Q_{23} = 6;$ \linebreak$Q_{24} = 4;$ $Q_{34} = 2;$ $Q_{35} = 4;$ $Q_{45} = 6;$ and\linebreak  $Q_{56} = 10.$  Updating $D_{ij}$ using $\Delta D_{ij} = \alpha |Q_{ij}| - r D_{ij}$, and taking   $\alpha = 1$, and  $r = 1$ one has that \begin{eqnarray}\Delta D_{12} &=& 8;~  \Delta D_{23} = 4;~  \Delta D_{24} = 2,\nonumber\\\Delta D_{34} &=& 0;~  \Delta D_{35} = 2;~ \Delta D_{45} = 4;~  \Delta_{56} = 8.\nonumber\end{eqnarray} 

The updated values of $D_{ij}$ then are: \begin{eqnarray}
D_{12} = 10;~D_{23} = 6;~ D_{24} = 4,\nonumber\\D_{34} = 2;~ D_{35} = 4;~ D_{45} = 6;~ D_{56} = 10.\nonumber\end{eqnarray}  After repeating this process several times,  a path  appears when a lot of the $D_{ij}$ are close to zero and then the rest are much greater.
\end{example}
\vspace{-0.1 in}
 \hrule

\subsection{Cellular model}\label{celu}

 In the cellular model as described in  \cite{plu8}, or {\it CELL model},  a cell is a mass of cytoplasm surrounded by a membrane, which is placed on a lattice grid where every square has an assigned number/state. State 2 represents cytoskeleton, state 1 represents cytoplasm, and state 0 represents squares that are not part of the CELL. The model is based on the following assumptions:
    \begin{enumerate}
    \item The membrane is the part of the cell where the cytoskeleton is concentrated and hardened;
    \item When a cytoskeleton assembly is taken apart and the membrane softens, the cytoplasm is distributed to other areas of the cell;
    \item Cytoplasmic flow is accompanied by transportation of the cytoskeleton and distribution of cytoskeleton assemblies.
    \end{enumerate}
    
    The CELL model has two phases: {\it development} and {\it foraging}.   In the {\it development phase}, the cell grows, forming a diamond shaped model with cytoskeleton around the edges and cytoplasm inside.  In the {\it foraging stage}, the model ``eats" zero which leads to a redistribution of cytoplasm and cytoskeleton/membrane. 
    The algorithm is as follows:
\begin{enumerate}
    	\item Chose a site with state 2, the {\it stimulus point};
    	\item Randomly choose a neighbor of the stimulus point in state 0 and replace the state of the stimulus point with the state of the chosen neighbor, so zero invades the cell. We call the zero a {\it bubble};
    	\item Replace state 1 with 2 so all non-zero cells are 2. Set number of moves to zero;
    	\item Mark the site with the bubble;
    	\item Decide whether $s$ sites of the bubble's neighbors are in state 0 or not. If yes, go to 8 otherwise go to 6;
    	\item Decide whether the number of moves exceeds threshold $n$. If yes, go to 8 otherwise go to 7;
    	\item Randomly choose one of the bubble's non-marked neighbors which is in state 2. Replace the state of the bubble with the state of chosen neighbor. Add 1 to number of moves, go to 4;
    	\item Reorganize the boundary and inside using algorithm described above. For example, if site with state 2 is surrounded by only neighbors with state 2, its state is changed to 1. Return to 1.
    \end{enumerate}
    
    It should be noted that depending on the location of the stimulation, the model behaves differently. If the stimulus point is chosen randomly, the CELL moves like an amoeba. If the stimulus is chosen from several active zones, an adaptive network with tentacles forms.

 \vspace{0.1 in}
 \hrule
 \begin{example}
   In order to illustrate the model, we shall consider the CELL model in a $5 \times 5$ grid. After the initial growth stage, the CELL model is seen in Figure \ref{box1} (a) below. By randomly choosing a cell (seen in grey) to be the stimulus point, and choosing a neighbor and swapping, one can see the following step in Figure \ref{box1} (b):

              \begin{figure}[H]
  \centering \includegraphics[scale = 0.08]{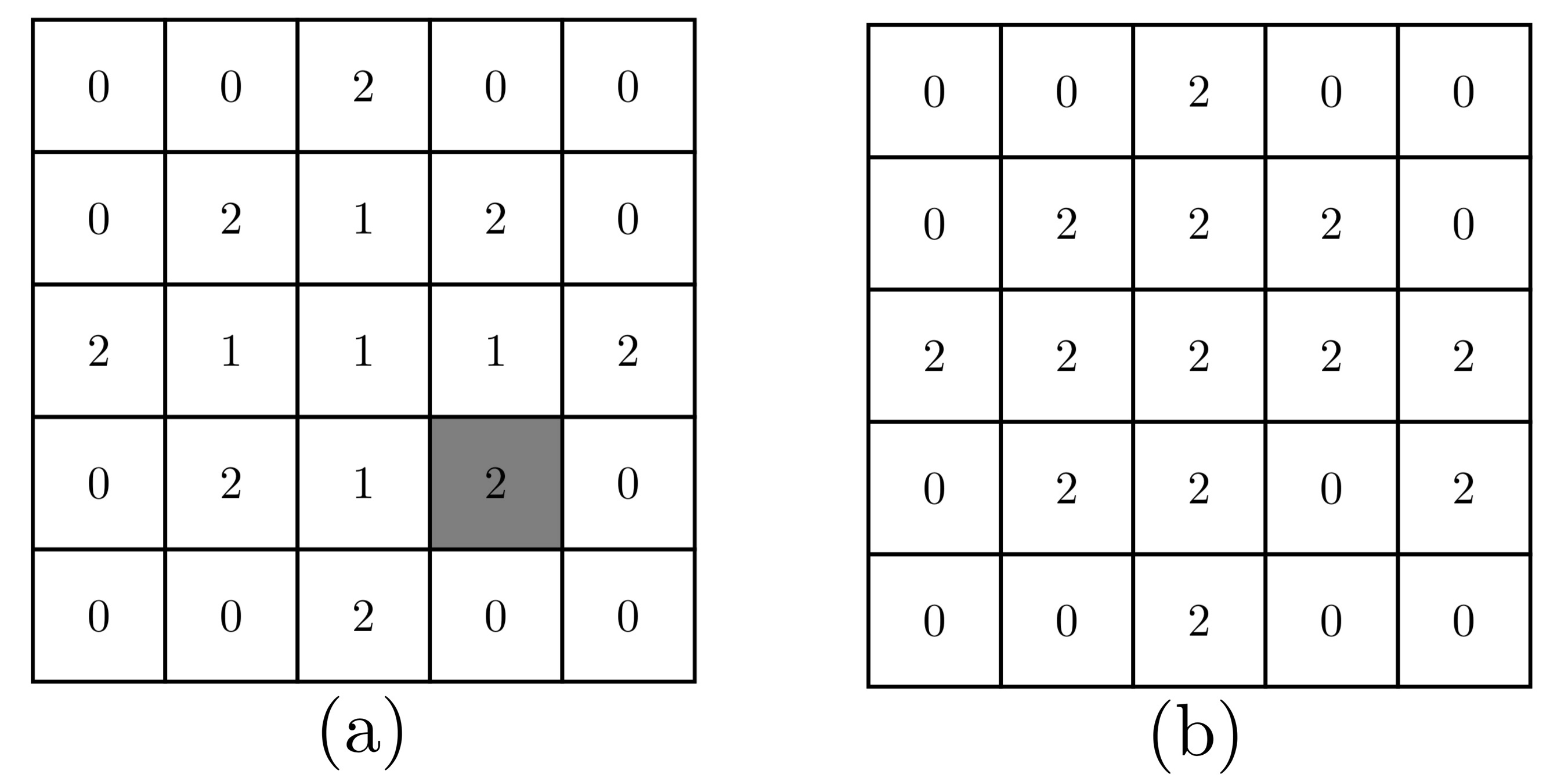}
  \caption{First two steps in a CELL model on a $5 \times 5$ grid.}
      \label{box1}
\end{figure}

In the following steps, one has the bubble randomly swap with neighbors more times. Doing this a couple more times one obtains Figure \ref{box2} (a). Finally, one can   reassign states to the cells, leading to Figure \ref{box2} (b).
              \begin{figure}[H]
  \centering \includegraphics[scale = 0.08]{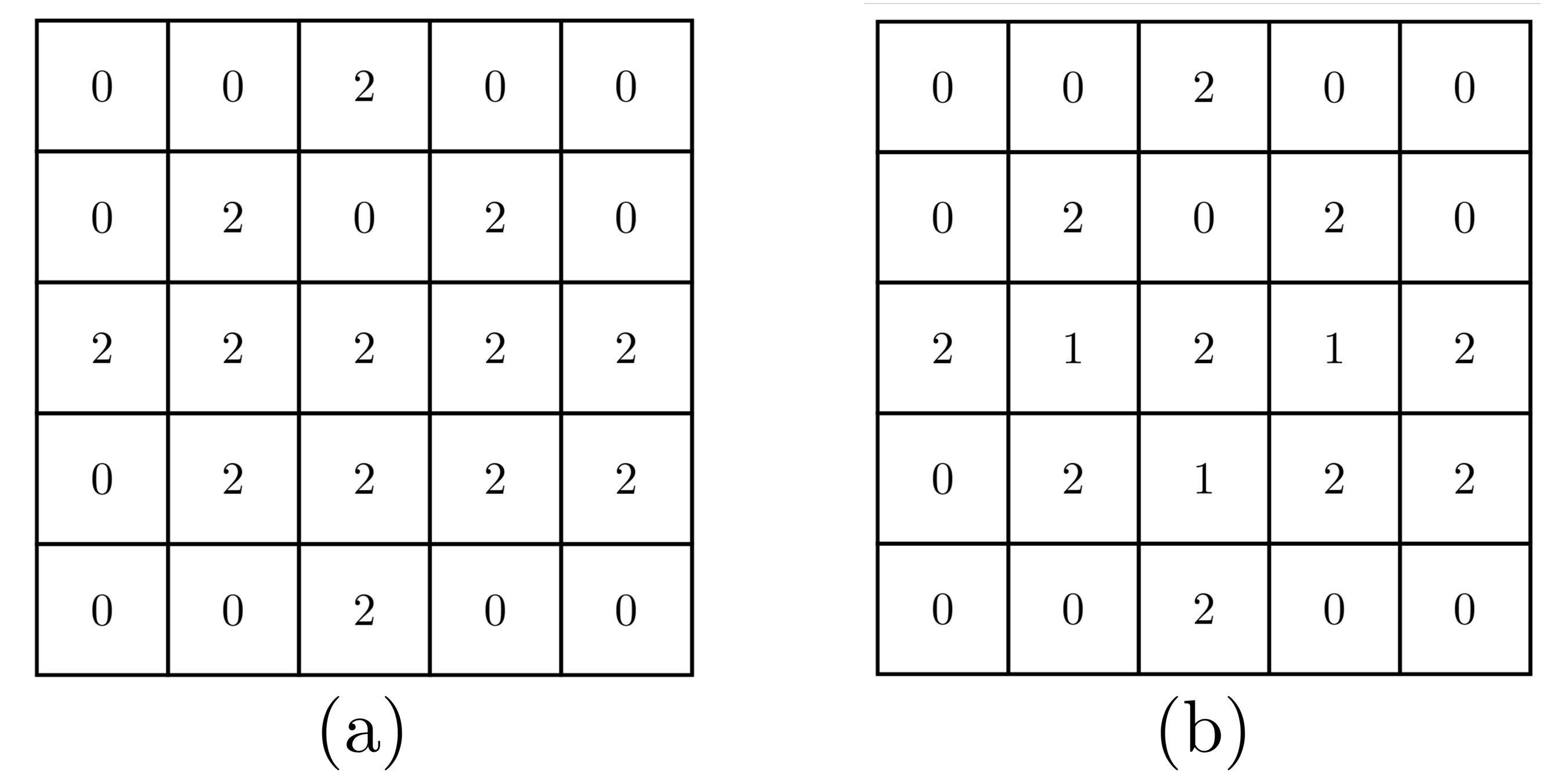}
  \caption{Last two steps in a CELL model on a $5 \times 5$ grid.}
      \label{box2}
\end{figure}
 \end{example}
 \vspace{-0.1 in}
 \hrule
  \vspace{0.1 in}

\begin{figure}[H]
 \centering \includegraphics[scale=0.35]{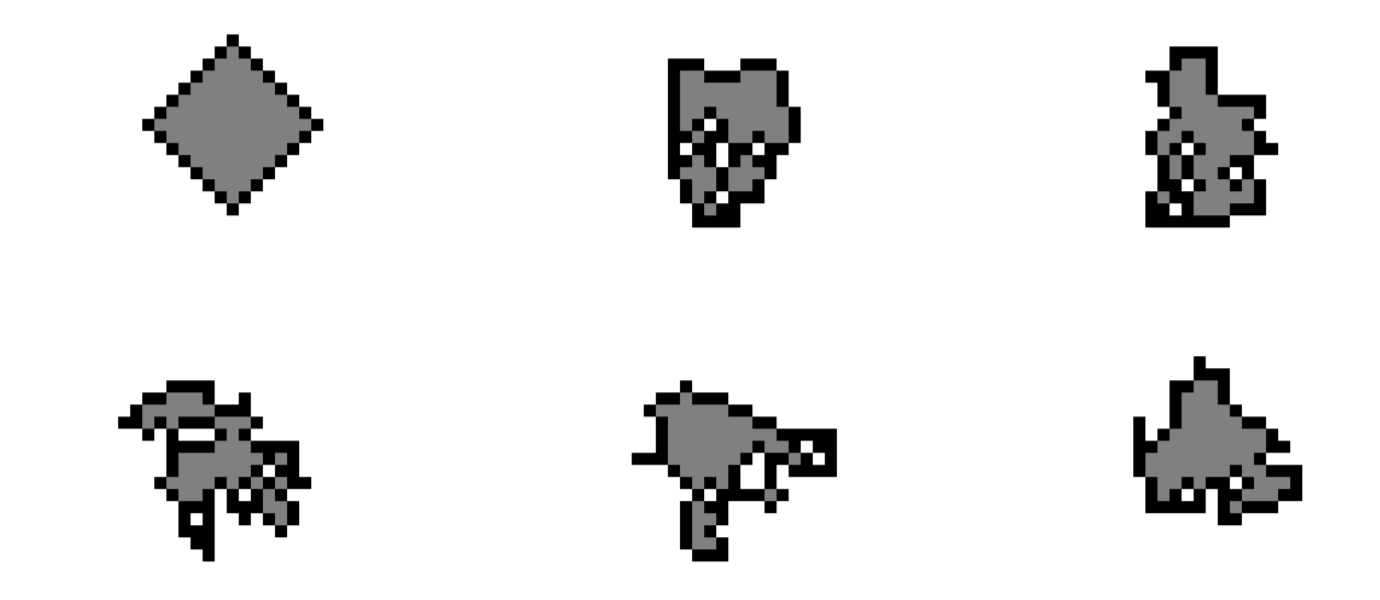}
 \caption{An example image from the implementation of the described CELL algorithm}
 \label{fig:example_cell}
 \end{figure}

\begin{figure}[H]
 \centering \includegraphics[scale=0.25]{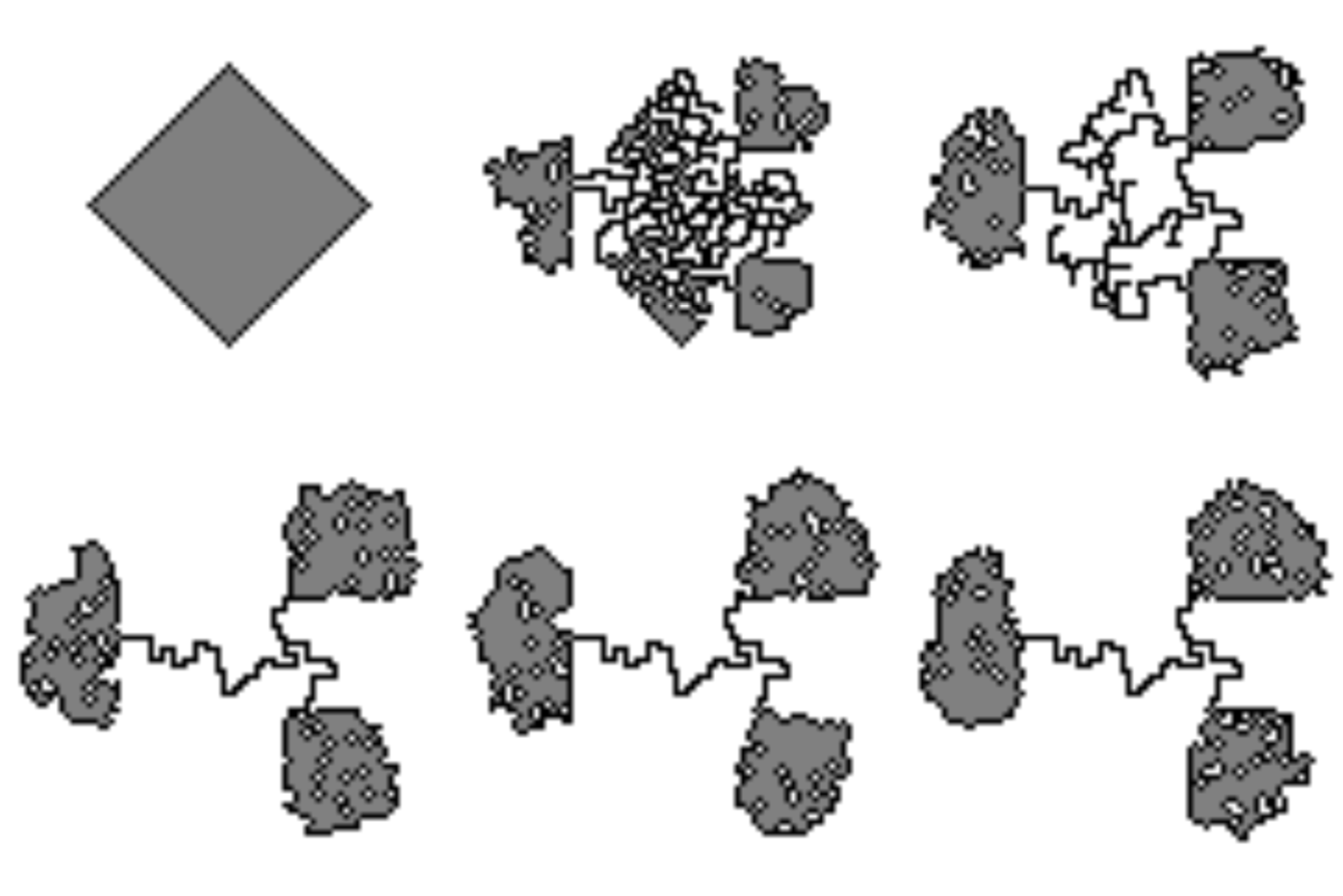}
 \caption{An example image from the implementation of the  CELL algorithm with active zones for tree formation}
 \label{fig:example_cell_network}
 \end{figure}

 A typical CELL  has bubbles within its cytoplasm. If enough of these bubbles group together, a chamber can form within the CELL. When this chamber bursts, tentacles are formed, as illustrated in Figure \ref{fig:example_cell}. 
 This model can also be used to form networks and find the shortest path. This is mainly done by creating active zones in which the stimulus point is always chosen from. For example, in Figure \ref{fig:example_cell_network}, active zones are created in three regions of the cell, leading to the cell forming a tree between the three active zones. Finally, this model has been previously used to solve mazes, generate spanning trees \cite{plu8}, and model crowd evacuation \cite{april1}.
 
\subsection{Multi-agent model}\label{multi}

Inspired by the many different behaviors \textit{Physarum} demonstrates, the authors of \cite{plu12} introduced a multi-agent model which consists of agents who travel throughout \textit{Physarum}, forming a chemoattractant map. We shall represent this as a large grid similar to the CELL model, and consider an agent to fill a cell of the grid. Each agent also has three sensors which sample values sensor offset distance (SO) away. The agents then uses the readings from the three sensors and orient themselves towards the strongest chemoattractant reading from the sensors. 
 
    There are two important parameters which need to be considered in this set up:
   \begin{itemize}
   \item There is sensor angle (SA), which is the angle between two of the three sensors;
   \item There is the rotation angle (RA) which is the amount the agent rotates to the strongest chemoattractant reading.
   \end{itemize}
    For example, if RA $<$ SA, contraction behavior is increased but if RA $>$ SA, spontaneous branching happens.
 
    The agent first reads from each of its 3 sensors. It turns by RA in the direction of the greatest sensor. After the agent has rotated accordingly, the agent attempts to move forward. If it can move forward (the cell isn't occupied), chemoattractant is deposited ( and the grid is updated) and then move. Agents are randomly chosen to rotate and then move. In addition, a $3 \times 3$ mean filter is applied on the grid to simulate diffusion. 
  Depending on where one initializes agents,  and what the SA and RA are, one gets different behaviors for the system, which can be summarized as follows:
 \begin{itemize}
 \item    {\it Filamentous condensation method.} Initialize a few agents at random locations and orientations. Then, a network forms. Add food sources which emit chemoattractants and the network will form a Steiner tree.
\item    {\it Filamentous Foraging Method.} Initialize agents at food sources.
 \item    {\it Plasmodial Shrinkage Method.} Randomly remove agents, other agents will move and form a minimum spanning tree.
\end{itemize}

\hrule 
\begin{example}
 To understand the multi agent model described above, we shall consider the following chemoattractant map shown in Figure \ref{box3} (a), and assume  one  has an agent with $SA$, and $RA = 45$ at the shaded square in Figure \ref{box3} (a). The rightmost sensor will then sense the highest chemoattractant (29) and the agent will rotate 45 degrees to the right. The sensor then moves forward if possible and deposits chemoattractant on the map (we shall assume it always deposits 5), and thus the next step in the model can be seen in Figure \ref{box3} (b) below.
 
                \begin{figure}[H]
  \centering \includegraphics[scale = 0.08]{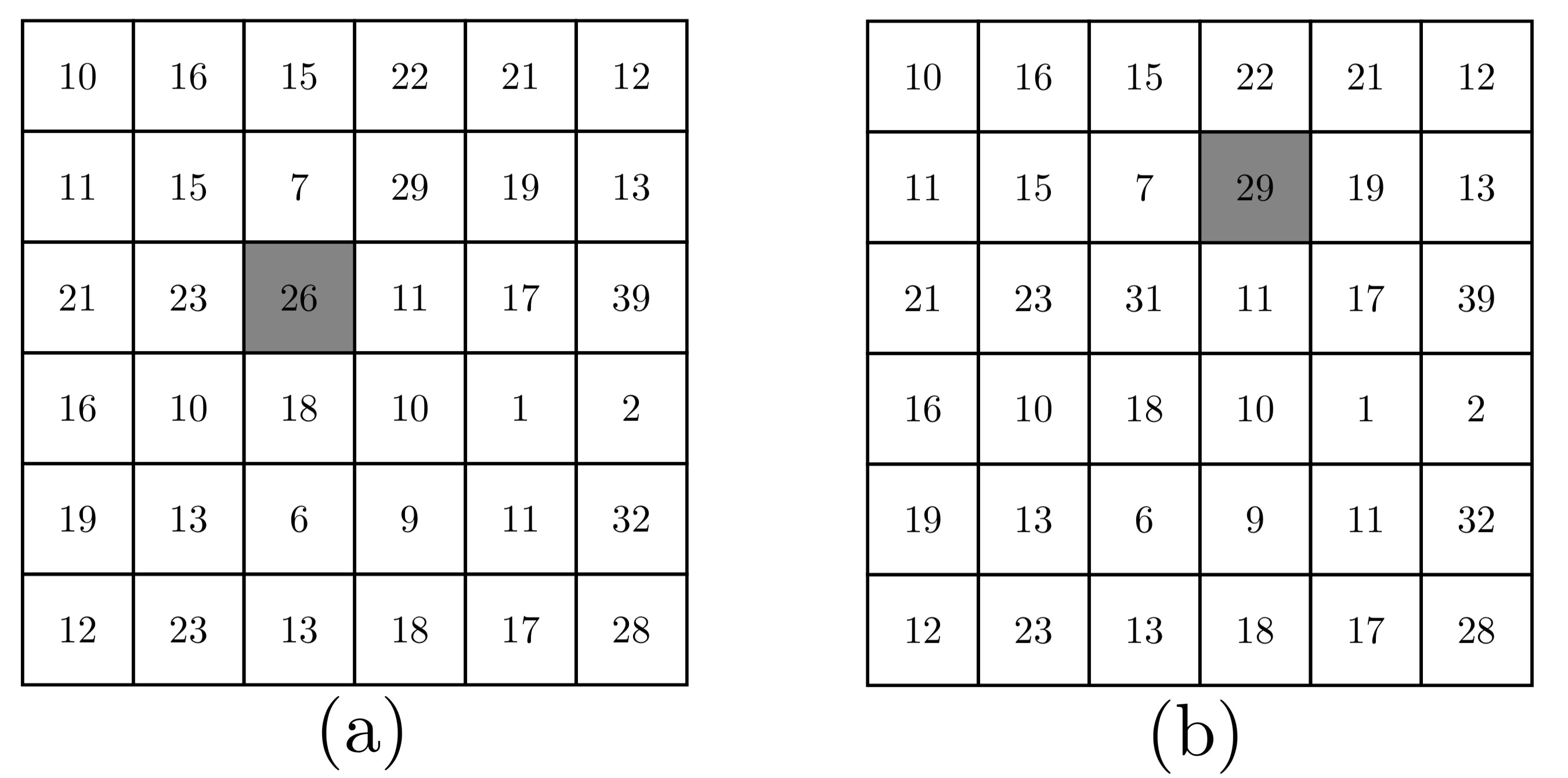}
  \caption{First two steps in a multi-agent model on a $6 \times 6$ grid.}
      \label{box3}
\end{figure}
  
 To simulate diffusion, we then apply a $3 \times 3$ mean filter to the map where we replace the value of each cell with the average of the cells in the $3 \times 3$ surrounding box, leading to the following Figure \ref{box4}.
     
                \begin{figure}[H]
  \centering \includegraphics[scale = 0.65]{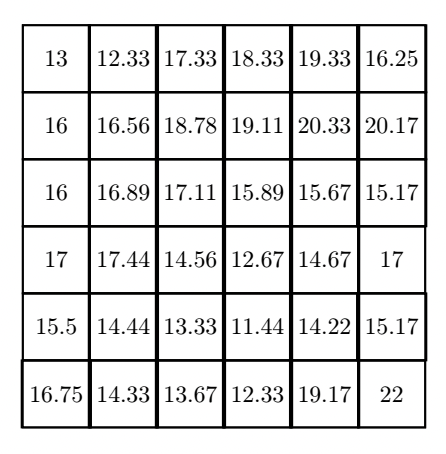}
  \caption{Grid with the  the average of the cells in the $3 \times 3$ surrounding box. }
      \label{box4}
\end{figure}
\end{example}

\hrule 
\vspace{0.1 in}

\begin{figure}[H]
  \centering \includegraphics[scale = 0.15]{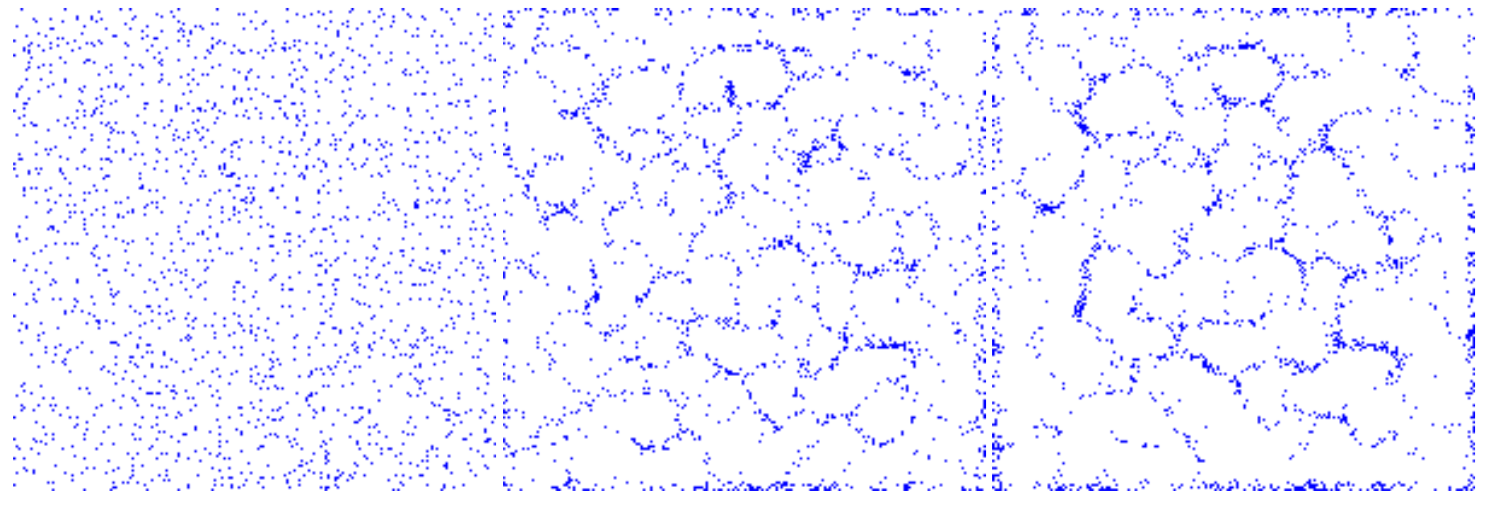}
  \caption{Images from a multi-agent model implementation. }
      \label{fig:example_multi_agent}
\end{figure}

Finally, it should be noted that through this model one can see  the agents gradually move from being randomly distributed to forming a network, as shown in  Figure \ref{fig:example_multi_agent}.
    
   \subsection{Shuttle-streaming model}\label{shut}
    
     We shall next describe the shuttle-streaming model following \cite{april2}. This model is based on three biological observations about \textit{Physarum polycephalum}:\begin{itemize}
     \item open ended tubes gradually disappear; \item when two tubes connect the same food sources, the longer one disappears; 
     \item outside changes like addition of nutrients can cause changes to the rhythmic contractions and shuttle streaming of \textit{Physarum}. \end{itemize}
      {\it Shuttle streaming} is the flow of protoplasm through \textit{Physarum}'s tubes and plays an important role in chemical signalling. When the organism contracts periodically (1-3 min), the direction of the shuttle streaming changes as hydrostatic pressure is produced.
 
In order to find the shortest path from node $v_S$ to $v_F$, which are two food sources, one considers two different protoplasmic flows, originating at each of the two nodes $v_S,$ and  $v_F$. 
 \bigskip
 
\noindent {\bf Forward flow ($v_S$ to $v_F$).}
The amount ${\rm Re}_f((v_i, v_j), t)$ of nutrients received from node $v_i$ through edge $(v_i, v_j)$ during forward flow is given by
\begin{equation}
{\rm Re}_f((v_i, v_j), t) = {\rm Se}_f((v_i, v_j), t) - \eta L_{ij},
\end{equation}
where ${\rm Se}((v_i, v_j), t)$ is the amount of nutrients sent through edge $(v_i, v_j)$ at node $v_i,$  the value $L_{ij}$ is length of edge, and $\eta$ is the absorption rate per unit length of tube wall. Additionally, ${\rm totRe}(v_j, t)$ is the total amount of nutrients that node $v_j$ receives.

The node $v_j$ distributes the total amount of nutrients received in forward flow proportionally to the amount received in previous backward flow. Hence, the amount of nutrients sent out through edge $(v_j, v_k)$ by node $v_j$ is given by 
\begin{eqnarray}
{\rm Se}_f((v_j, v_k), t) &=& {\rm totRe}_f(v_j, t) \times \frac{{\rm Re}_b((v_k, v_j), t - \delta t)}{{\rm totRe}_b(v_j, t - \delta t)}\nonumber
\\
{\rm totRe}_f(v_j, t) v &=&\sum_{\forall v_i \in (v_i, v_j)} {\rm Re}_f((v_i, v_j), t)\nonumber
\\
{\rm totRe}_b(v_j, t - \delta t) &=&\sum_{\forall v_k \in (v_k, v_j)} {\rm Re}_b((v_k, v_j), t - \delta t)\nonumber
\end{eqnarray}
 \bigskip
 
\noindent  {\bf Backwards flow at time $t + \delta t$}.
    Similar to the forward flow, one has that
    \begin{equation}
    	{\rm Re}_b((v_p, v_q), t + \delta t) = {\rm Se}_b((v_p, v_q), t + \delta t) - \eta L_{pq}.\nonumber
    \end{equation}
The corresponding equations for 
\begin{eqnarray}
\phi:=\frac{{\rm Re}_f((v_r, v_q), t)}{{\rm totRe}_f(v_q, t)}\nonumber
\end{eqnarray}
are given by 
   \begin{eqnarray}
{\rm Se}_b((v_q, v_r), t + \delta t) &=& {\rm totRe}_b(v_q, t + \delta t) \times \phi;\nonumber
\\
{\rm totRe}_b(v_q, t + \delta t) &=& \sum_{\forall v_p \in (v_p, v_q)} {\rm Re}_b((v_p, v_q), t + \delta t);\nonumber
\\
{\rm totRe}_f(v_q, t  ) &=& \sum_{\forall v_r \in (v_r, v_q)} {\rm Re}_f((v_r, v_q), t).\nonumber
\end{eqnarray}
 
 The two  nodes $v_S$ and $v_F$ act as both source and sink nodes depending on the direction of the protoplasmic flow. Assuming that the amount of nutrients the tubular network absorbs per one forward or backward flow session is $N_{in}$,   the amount of nutrients sent and received by nodes $v_s, v_F$ can be computed in the following fashion. 
 
 At $t = 0$ one can compute the number of nutrients sent from node $v_j$ as follows:
 \begin{equation}
 {\rm Se}_f((v_j, v_k), T = 0) = \frac{{\rm totRe}_f(v_j, t = 0)}{\sum\limits_{\forall v_k \in (v_j, v_k)}1}
 \label{eq:Sef}
 \end{equation}
    This can be thought of as the total amount of nutrients divided by the number of edges. When the nutrients carried by an edge is zero or negative, we can discard the edge since biologically it would die without nutrients. One continues to do this until  the end of the optimization process is reached, at which point each node only has one in-edge and one out-edge.

  \vspace{0.1 in}
 \hrule
\vspace{0.1 in}
\begin{example}
To illustrate this model, consider a similar graph to the flow-conductivity model of Figure \ref{col2}:
   \begin{figure}[H]
  \centering \includegraphics[scale = 0.3]{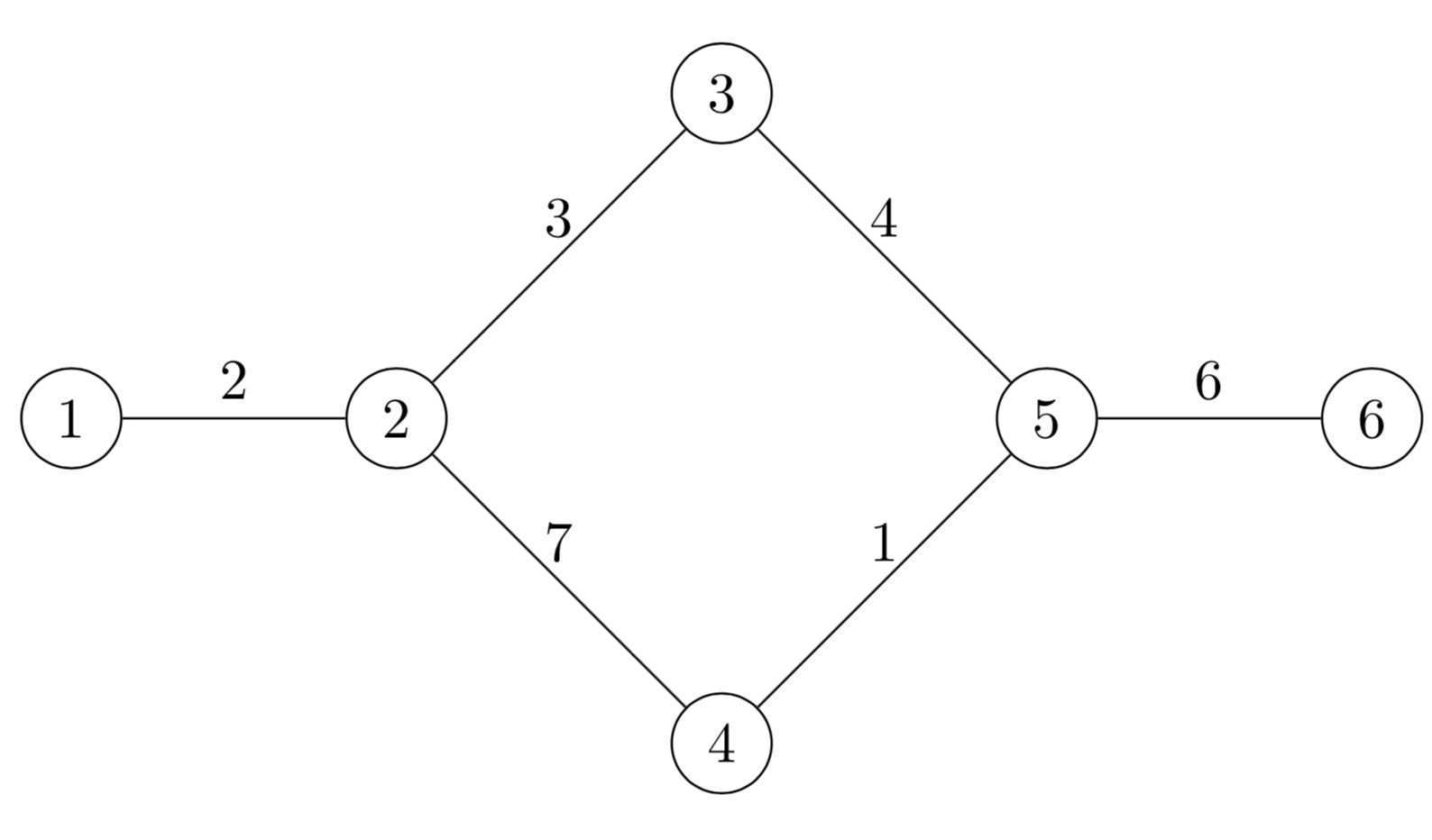}
  \caption{Graph as in Figure \ref{col2} but without the edge between nodes 3 and 4. The numbers on each edge represent the length of that edge.}
      \label{col3}
\end{figure}
\pagebreak
 
Let $N_{in} = 10$, which means one puts in 10 nutrients every time. The starting values can be then computed using Equation \eqref{eq:Sef}. Let $\eta = 0.1$. Then we have that
 
 \begin{eqnarray}{\rm Se}_f((1, 2), 0)  &=& 10,\nonumber\\ {\rm Re}_f((1,2), 0)  &=& 9.8, \nonumber\\{\rm Se}_f((2, 3), 0)  &=& 4.9, \nonumber\\{\rm Re}_f((2, 3), 0)  &=& 4.6, \nonumber\\{\rm Se}_f((2, 4), 0)  &=& 4.9, \nonumber\\{\rm Re}_f((2, 4), 0)  &=& 4.2, \nonumber\\{\rm Se}_f((3, 5), 0)  &=& 4.6, \nonumber\\{\rm Re}_f((3, 5), 0)  &=& 4.2, \nonumber\\{\rm Se}_f((4, 5), 0)  &=& 4.2, \nonumber\\{\rm Re}_f((4, 5), 0)  &=&4.1, \nonumber\\{\rm Se}_f((5, 6), 0)  &=& 8.3, \nonumber\\{\rm Re}_f(5, 6)  &=& 7.7.\nonumber\end{eqnarray}

The values of ${\rm totRe}_f(v, 0)$ can be computed in a similar manner,  and are given by
 \begin{eqnarray} {\rm totRe}_f(1, 0)  &=& 10,  \nonumber\\{\rm totRe}_f(2, 0) &=& 9.8,  \nonumber\\{\rm totRe}_f(3, 0)  &=& 4.6,  \nonumber\\{\rm totRe}_f(4, 0)  &=& 4.2, \nonumber\\ {\rm totRe}_f(5, 0)  &=& 8.3, \nonumber\\ {\rm totRe}_f(6, 0)  &=& 7.7.\nonumber\end{eqnarray}  
 
Now it is time for backwards flow. Starting from the opposite end, one sets ${\rm Se}_b((6, 5), 1) = 10$, and so ${\rm Re}_b((6, 5), 1) = 9.4.$ Then, one has that 
\begin{eqnarray}{\rm Se}_b((5, 3), 1) &=& 4.7566267,\nonumber\\ {\rm Re}_b((5, 3), 1) &=& 4.3566267,\nonumber\\
 {\rm Se}_b((5, 4), 1) &=& 4.64337, \nonumber\\, {\rm Re}_b((5, 4), 1) &=& 4.54337,\nonumber\\ {\rm Se}_b((3, 2), 1) &=&  4.3566267, \nonumber\\ {\rm Re}_b((3, 2), 1) &=& 4.0566267,\nonumber\\ {\rm Se}_b((4, 2), 1) &=& 4.54337, \nonumber\\  {\rm Re}_b((4, 2), 1)&=& 3.84337,\nonumber\\ {\rm Se}_b((2, 1), 1) &=& 7.9, \nonumber\\ {\rm Re}_b((2, 1), 1) &=& 7.7.\nonumber\end{eqnarray}  As done with forward flow, one can compute ${\rm totRe}_b$ and then repeat the same procedure with forward flow. After repeating the procedure several times, the flow through certain edges will eventually become very small at which point one can remove such edges until obtaining  the shortest path. 
\end{example}
 \vspace{0.1 in}
 \hrule
 \vspace{0.1 in}

Since the model is   a shortest path algorithm, it can be thought of as being very similar to flow-conductivity, so any of the shortest path applications such as traffic routing should be possible, although this model has been shown to not always give the shortest path.

\section{Exploring interactions between multiple cells}\label{content}
In what follows we shall introduce a modified version of  the CELL model  described theoretically in \cite{plu8} with memorized flow. After creating code to model our new algorithm, we shall investigate the interactions between multiple cells. To do this, our new model incorporates a unique ID corresponding to a color for each cell, so that each cell is spawned with a unique   color, allowing us to distinguish between the cytoplasm of each cell. 
 \vspace{0.05 in}
\hrule
\begin{example}
An example of  our model  is shown in Figure \ref{fig:twoSmallCellsCombine}, where one can see two CELLs initially spawned in a diamond shape. After a few iterations, one can see the cells lose their diamond shape and become more circular, and there are multiple chambers within the cells.
\begin{figure}[H]
\centering\includegraphics[scale=0.2]{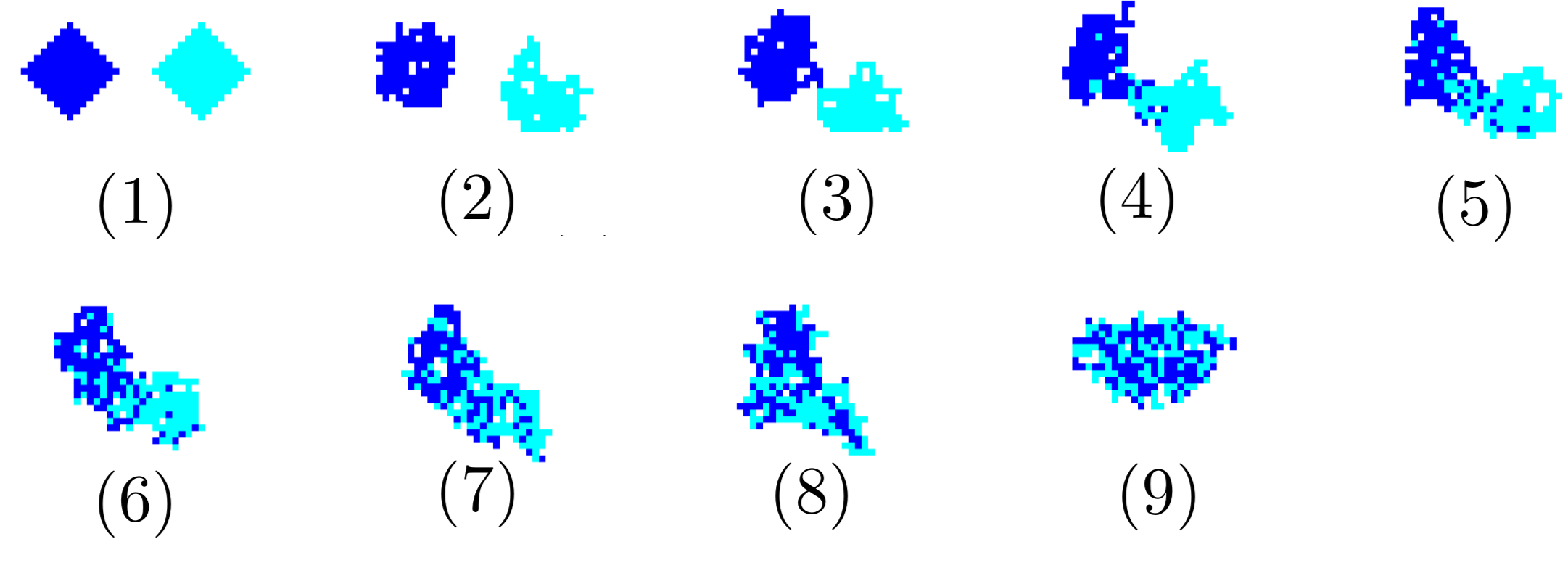}
 \caption{The spawn of two CELLs at (15, 20) and (35, 20), with parameters: size = 15, s = 3, n = 1000.}
 \label{fig:twoSmallCellsCombine}
 \end{figure}
 By the third image of Figure \ref{fig:twoSmallCellsCombine}, the cells have made contact. In the fourth image, cytoplasm continues to flow between the two newly connected cells. In the fifth and sixth image, cytoplasm continues to flow further between the two cells. Finally in the seventh image, the cytoplasm appears to start becoming well mixed and in the eight image, the cell seems to have taken on a more circular or compact shape after initially being elongated due to two cells combining. In the ninth image, the cell seems circular and the cytoplasm is well dispersed. In the following sections we shall see further analysis of how cells fuse through our model. 
 \end{example}
 \hrule
\vspace{0.1 in}

%In this section, we will fit curves to linear  \eqref{eq:linear}, exponential \eqref{eq:exponential}, and logistical \eqref{eq:logistical} functions.
%\begin{equation}
%y = mx + b
%\label{eq:linear}
%\end{equation}
%\begin{equation}
%y = a \cdot b^x + c
%\label{eq:exponential}
%\end{equation}
%\begin{equation}
%y = \frac{L}{1+e^{-k (x - x_0)}}+b
%\label{eq:logistical}
%\end{equation}

 \subsection{Time to cell fusion in relation to spawn distance}
\begin{figure}[H]
\centering
 \includegraphics[scale=0.05]{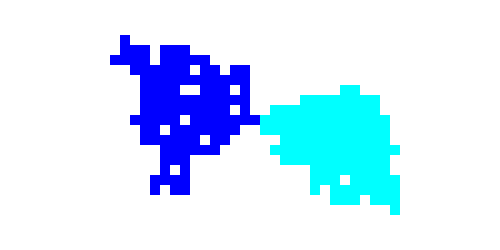}  
  \caption{In this image, the two cells have just contacted for the first time.}
  \label{fig:firstConnectionExample}
\end{figure}

\begin{figure}[H]
\begin{center}
\includegraphics[scale=0.13]{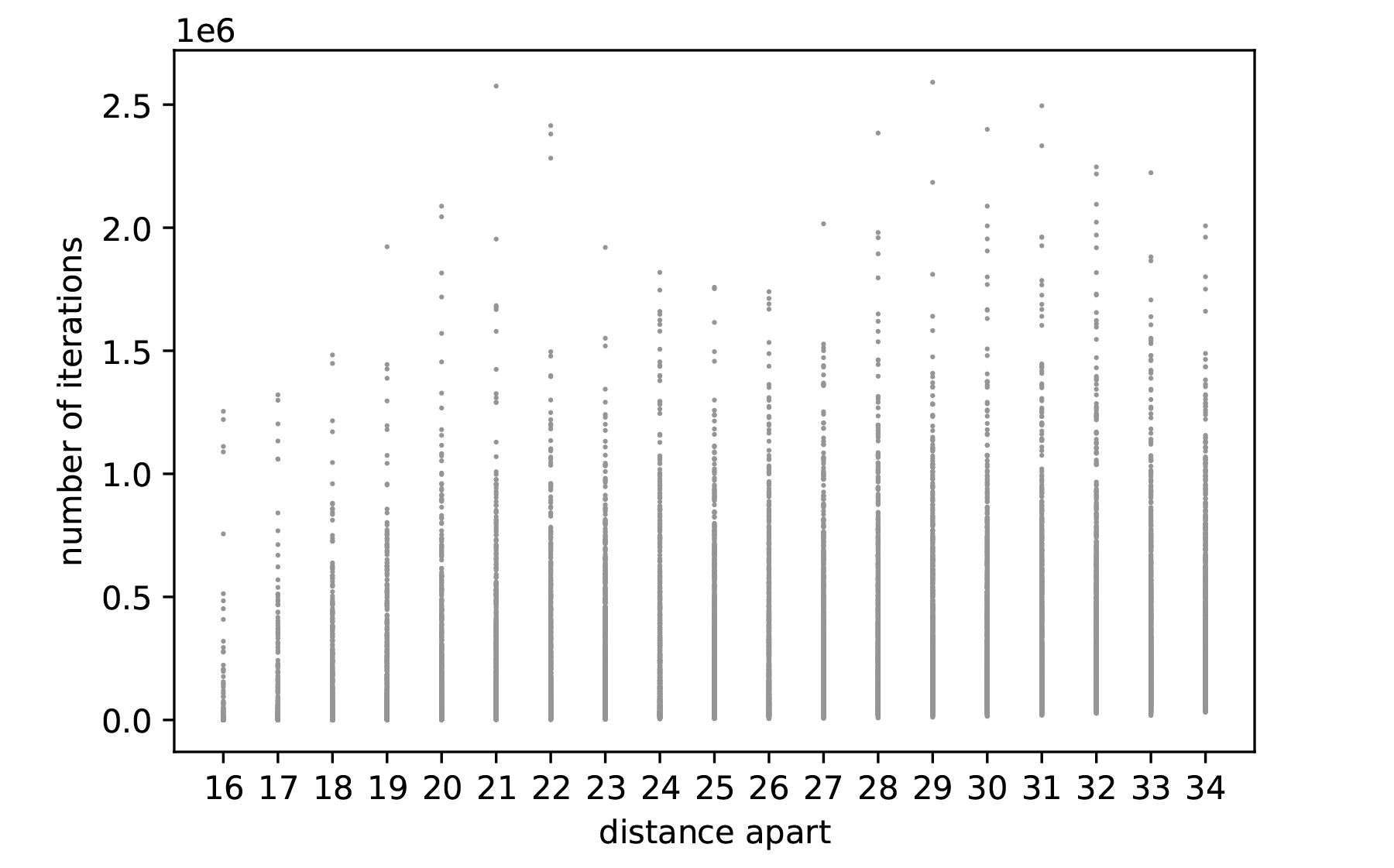}
\caption {Graph of the 1025 trials of each distance: the distance between the center of the spawn points is on the $x$ axis, and the number of iterations until they merge on the $y$ axis.}
\label{fig:scatterplot_mergedis}
\end{center}
\end{figure}

\begin{figure}[H]
\begin{center}
\includegraphics[scale=0.12]{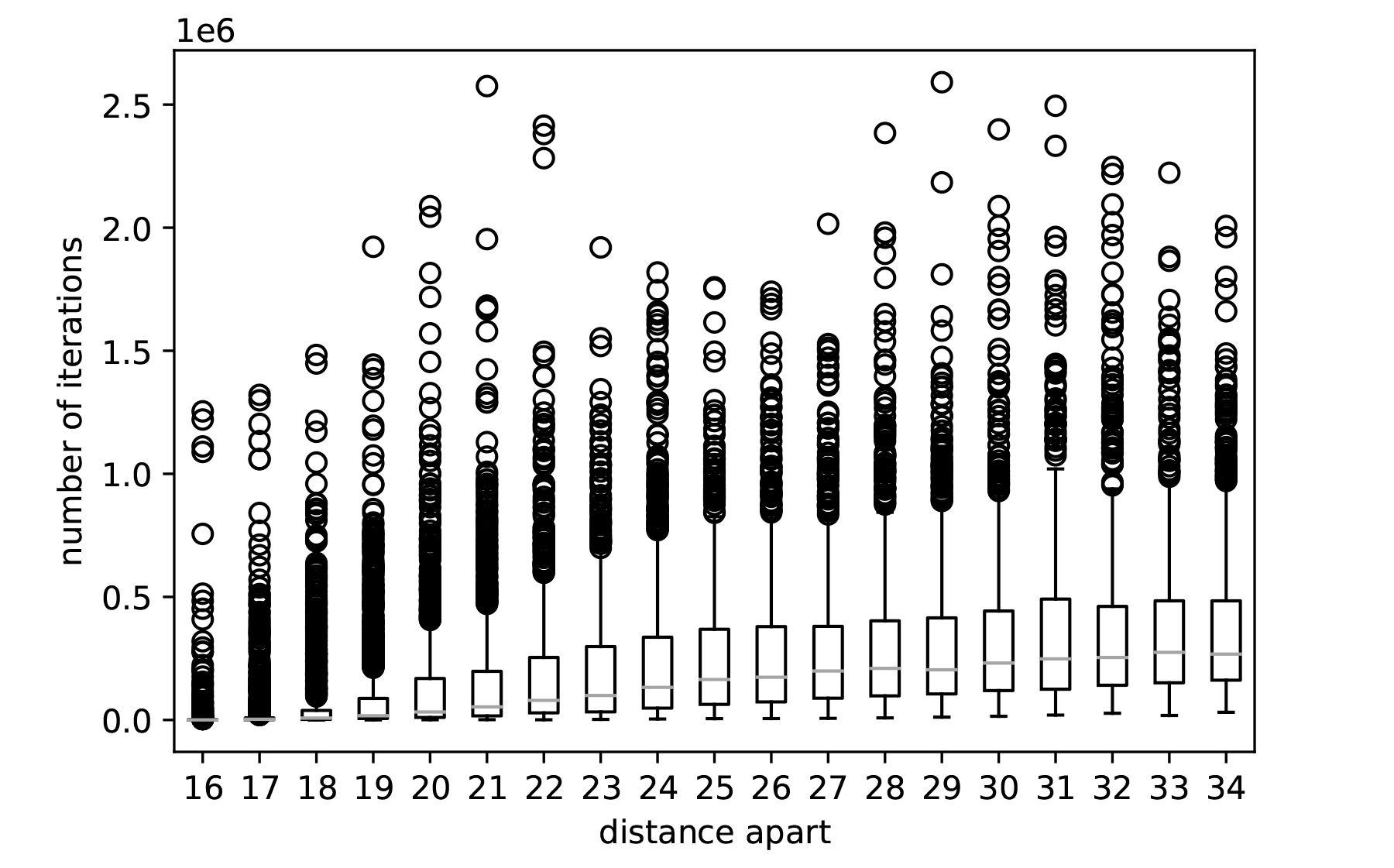}
\caption{The 1025 trials of each distance: the distance between the center of the spawn points is on the $x$ axis and the number of iterations until they merge on the $y$ axis.}
\label{fig:boxplot_mergedis}
\end{center}
\end{figure}

We shall dedicate this section to understanding how the distance two CELLs are spawned apart affects the number of iterations it takes for the cells to contact. Across the section, we use the following parameters: n = 1000, s = 3, and define two CELLs contacting as a cytoplasm of one ID being adjacent to a cytoplasm of another CELL. To illustrate this definition, Figure \ref{fig:firstConnectionExample} is an example of the first place two CELLs contact.
To understand the behavior of two cells through our model,  we run 1025 trials of each spawn distances from 16 to 34. 
We graph the average iterations of each trial versus the spawn distance in Figures \ref{fig:scatterplot_mergedis} and \ref{fig:boxplot_mergedis}. These plots shows that there are extreme upper outlines, which is expected given the random nature of the cell model. The boxplot also shows an overall increasing trend in number of iterations as the spawn distance increases.

In order to understand the results inferred from our experiments, we  fit our findings to curves to linear, exponential, and logistical functions:

\begin{eqnarray}
y &=&mx + b
\label{eq:linear}
\\
y &=& a \cdot b^x + c
\label{eq:exponential}
\\
y &=& \frac{L}{1+e^{-k (x - x_0)}}+b.
\label{eq:logistical}
\end{eqnarray}

\begin{figure}[H]
\begin{center}
\includegraphics[scale=0.6]{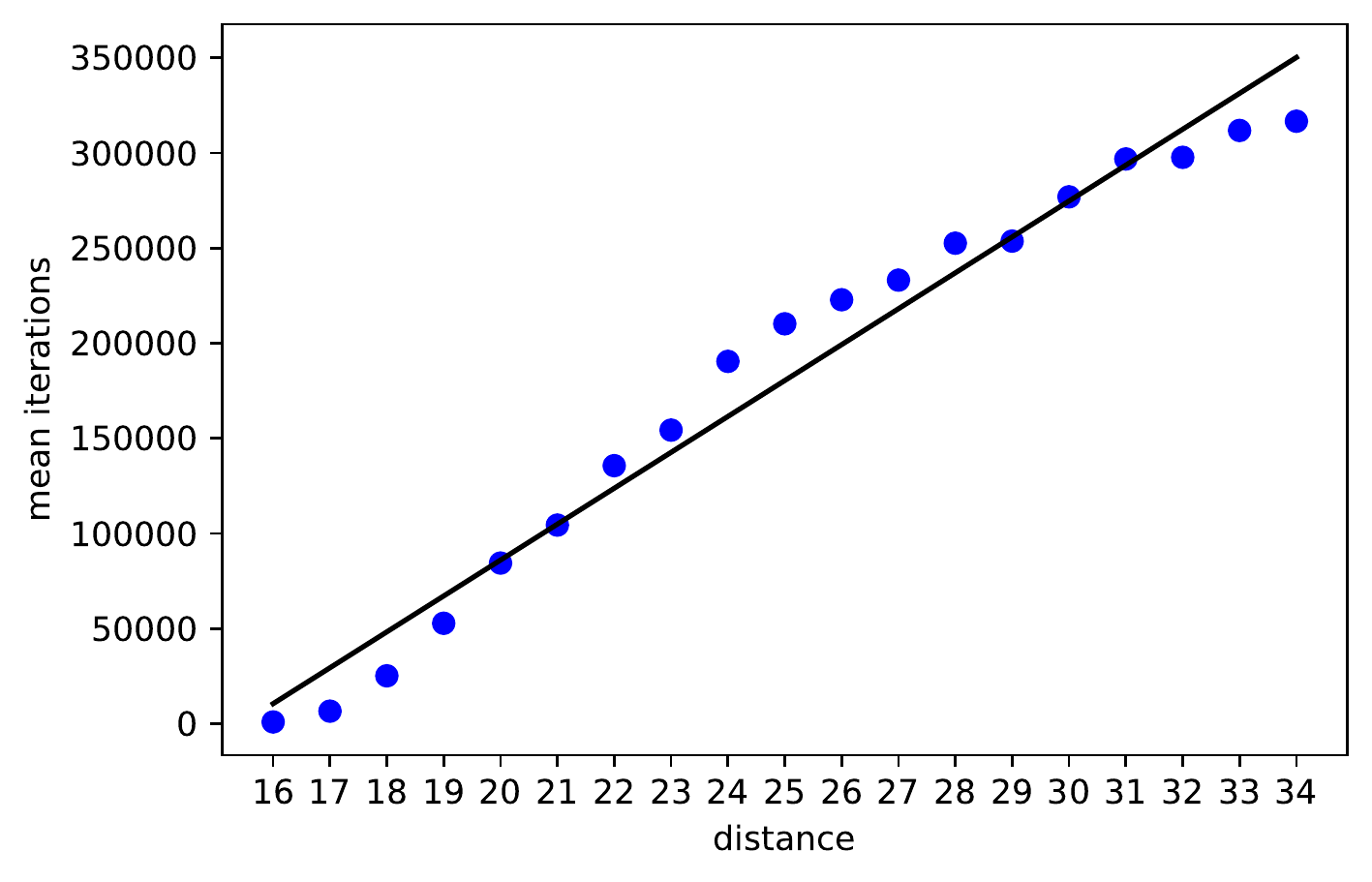}

\caption{To deal with the heavily skewed data as evidenced by the scatter and box plots, we trim 10 percent from both the upper and lower ends of the data. We then compute the trimmed mean. We use the scipy.optimize.curve\_fit function to fit this data to a linear function (Eq.~\eqref{eq:linear}) with the values $m = 21892.74674907$ and $b =- 354967.6942326$. This gives a $r^2$ value of $0.970675948980436$. }
\label{fig:linfit_mergedis}
\end{center}
\end{figure}

  The 10 percent trimmed mean of each spawn distance is given in Figure \ref{fig:linfit_mergedis}, where one can see that this data appears to be linear (Eq.~\eqref{eq:linear}) with values of $m=18873.73644734$ and $b=-291495.3914131$. It can be seen that the linear function gives an $r^2$ value of $0.9707$ and  appears to fit the data, allowing us to  conclude that the relationship between time to merge and the distance apart is linear.
%\begin{equation}
%y = 21892.74674907x - 354967.6942326
%\label{eq:fit_mergedis}
%\end{equation}

\subsection{Analysis of cell fusion with three cells}
\begin{figure}[H]
\begin{center}
\includegraphics[scale=0.09]{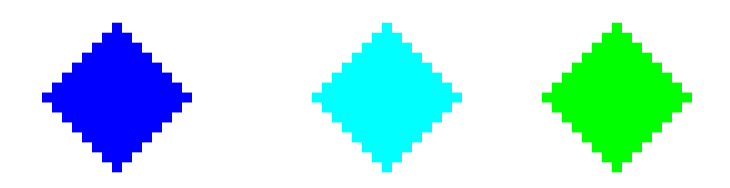}
\caption{Three cells of size 15,  labeled from left to right as 1, 2, and 3. The distance between cells 1 and 2 ({\it distance 1}) is 27 and between cells 2 and 3 ({\it distance 2}) is 23.}
\label{fig:sample_three}
\end{center}
\end{figure}

The premise of this experiment is that we spawn three CELLs in a straight line. We adjust the distances between the cells in order to see which cells fuse first as well as gather data on the number of iterations taken for the first cell fusion event to occur. We try different distances between the three CELLs ranging from 16 to 29 inclusive. We keep track of how many iterations it takes for the first two CELLs to merge and which two cells end up merging. We perform 350 trials. In general, the distance between cells 1 and 2, or {\it distance 1},  will be greater than or equal to the distance between cells 2 and 3, or {\it distance 2}. We always spawn cell 2 in the center of the image and then spawn cells 1 and 3 the set distance away from cell 2. For example, in Figure \ref{fig:sample_three}, the cells are from left to right 1, 2, and 3. 

\begin{figure}[H]
\includegraphics[scale=0.65]{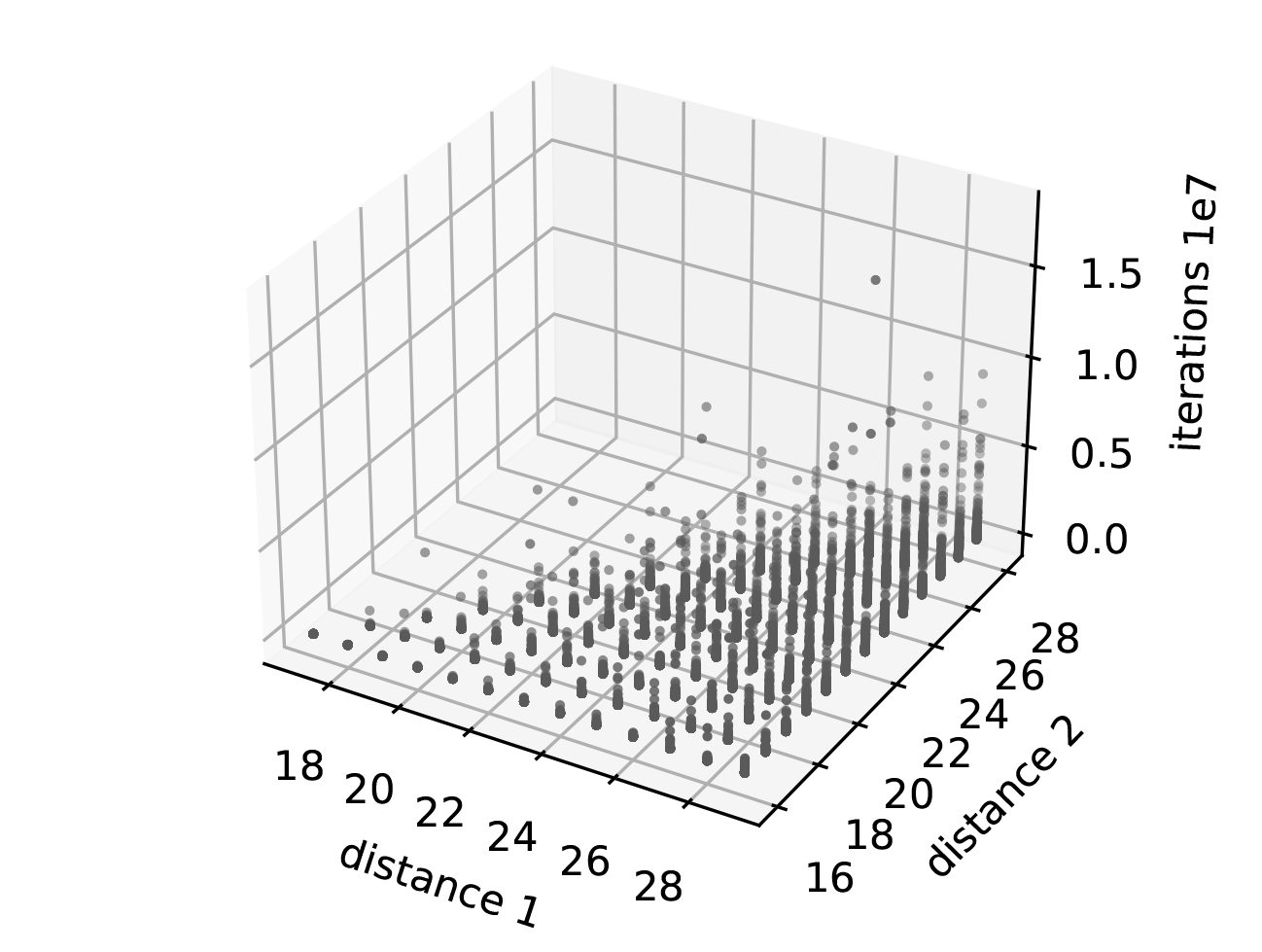}
\caption{Using the python package matplotlib, we create a 3D scatterplot. On the $xy$ plane are the distances between the three cells and on the $z$ axis is the number of iterations it takes for two cells to combine.}
\label{fig:3D_scatter_plot}
\end{figure}

In order to visualize the data from our experiments, we use a a 3D scatterplot in 
Figure \ref{fig:3D_scatter_plot}, through which  we see that there are extreme outlines  when both distances are very large.

\begin{figure}[H]
\includegraphics[scale=0.6]{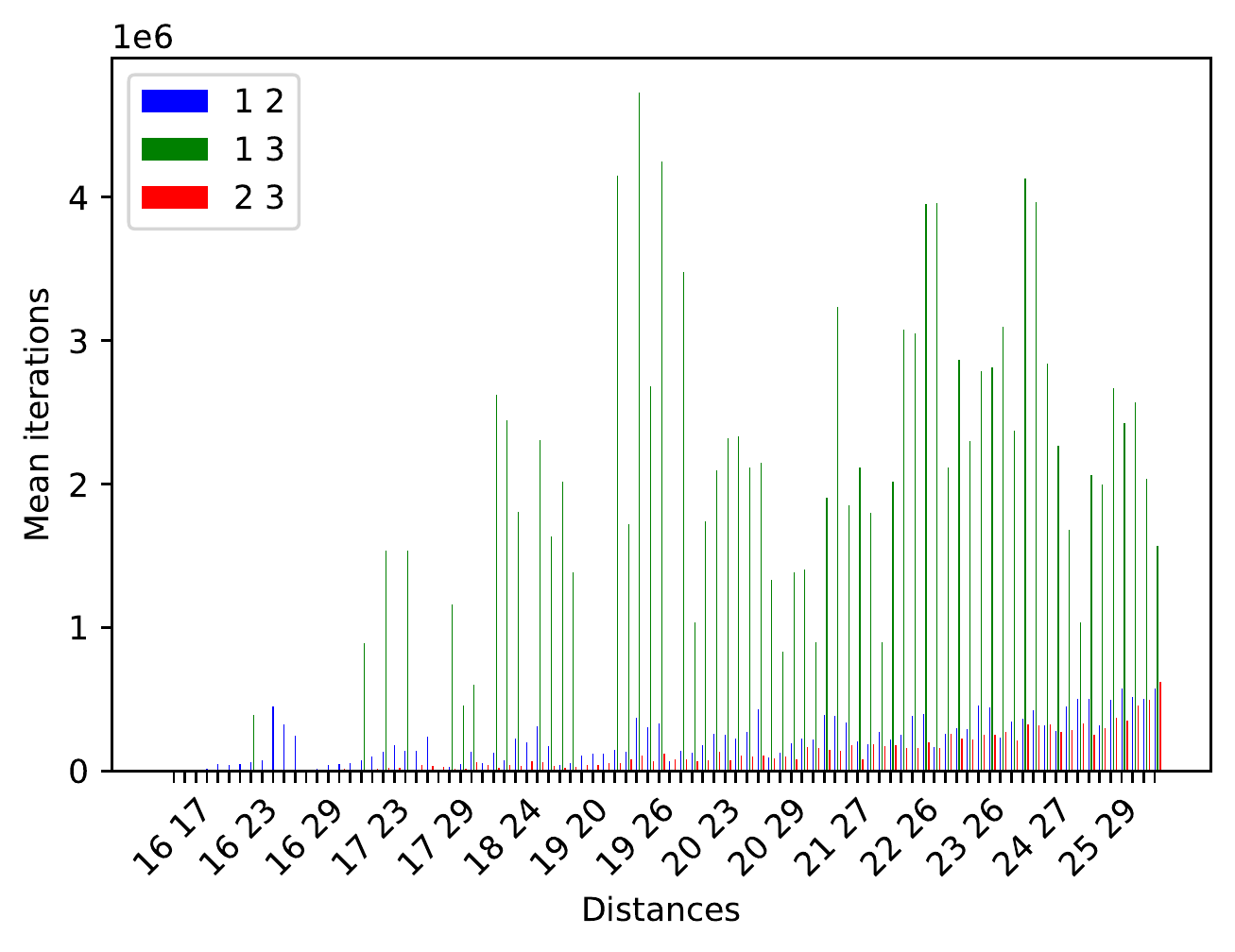}
\caption{Using the python package matplotlib, we compute the mean of each set of distances. For each distance, we have three bars: 1 for each of the possible two CELLs combining.}
\label{fig:bar_graph_mean}
\end{figure}

Moreover, one can see that for every set of distances, there are 3 different possibilities: cells 1 and 2 join, cells 2 and 3 join, or cells 1 and 3 join. For each of these possibilities, we compute the mean number of iterations it took for the cells to fuse. This is displayed in Figure \ref{fig:bar_graph_mean}. One sees that as the distances increase, so does the number of iterations. In general, it appears that the number of iterations for cells 1 and 3 (green bar) to join is overall much larger than the number of iterations for 1 and 2 or 2 and 3. 

\begin{figure}[H]
\includegraphics[scale=0.6]{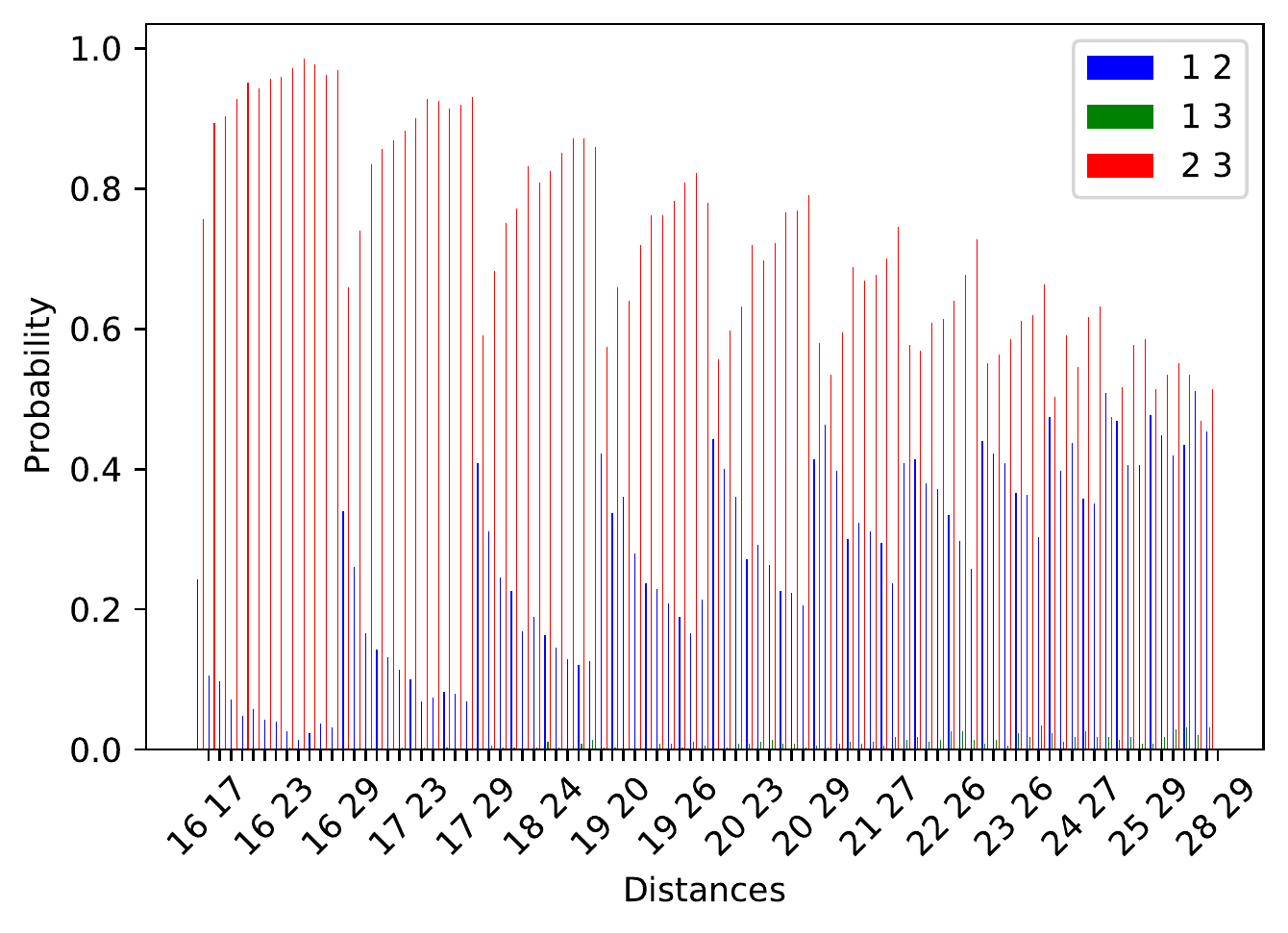}
\caption{Using the python package matplotlib, we compute the probability that for a given set of distances between CELLs, each combination of two cells (1 2,  1 3, 2 3) will be the ones that combine first.}
\label{fig:prob_bar}
\end{figure}

In order to understand how and when two cells fuse together,  we compute the probability that a certain two cells will join, depicted in Figure \ref{fig:prob_bar}. Overall, one sees that the probability cells 1 and 3 join is the lowest, which is expected since cell 2 is between cells 1 and 3. Cells 2 and 3 have the highest probability of joining which is consistent with the fact that they are closer.

\begin{figure}[H]
\includegraphics[scale=0.25]{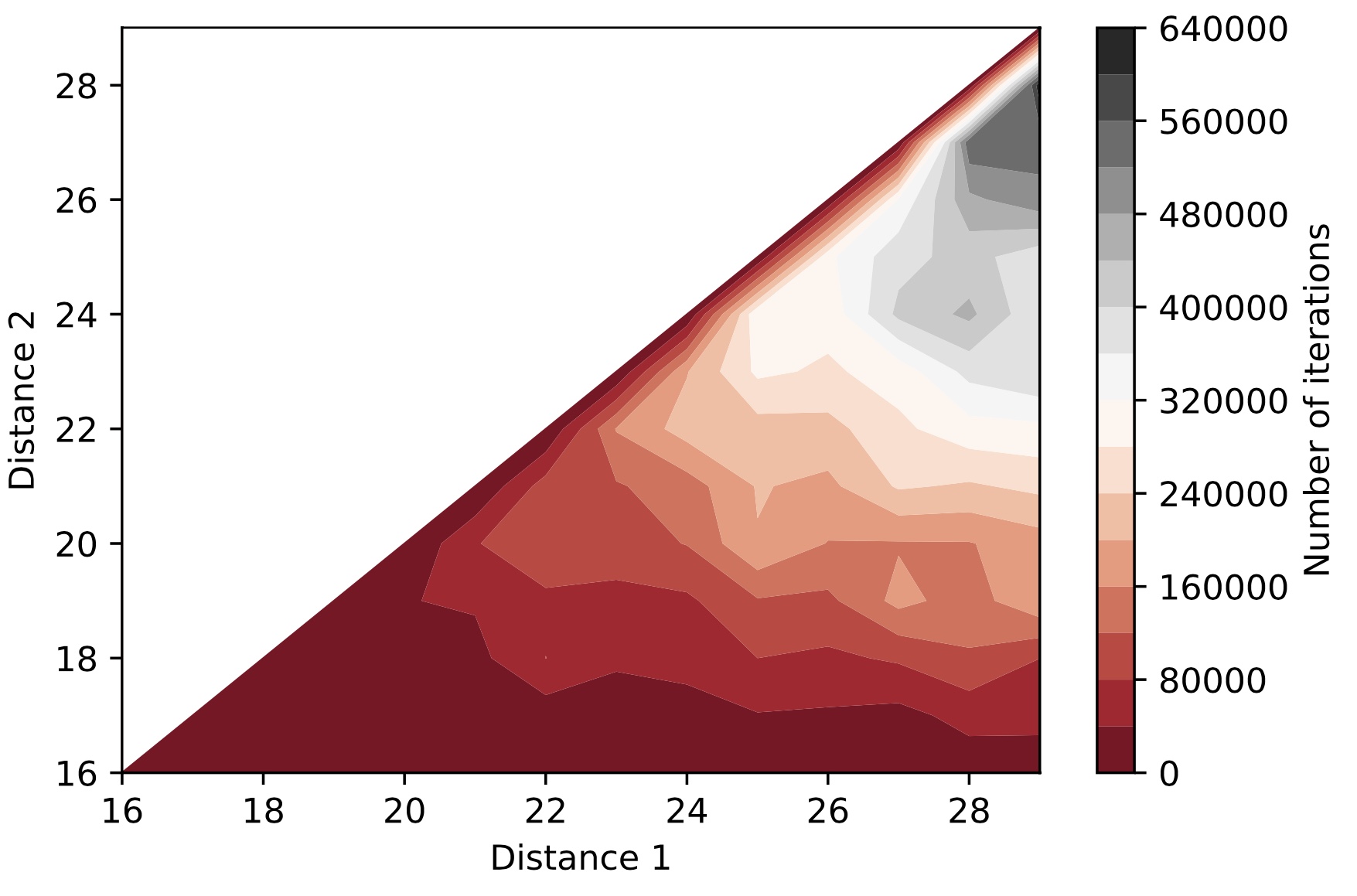}
\caption{Using the python package matplotlib, we create a contour plot. On the xy plane are the distances between the three cells and represented by the color gradient is the number of iterations it takes for two cells to combine.}
\label{fig:contour_plot}
\end{figure}

The number of iterations it takes for two out of the three cells to fuse steeply increases as the distance between the cells does, and this is shown in Figure \ref{fig:contour_plot}. It is important to note that distance 1 was always greater than distance 2, so the upper triangle of this plot had no data. 

\subsection{The impact of cell size on movement}
In what follows we shall show the importance of the CELL's size in its ability to move. In this section, we measure mobility by computing three values: the average distance from the spawn point, the furthest distance from the spawn point, and the total distance traveled. After every iteration of the cell algorithm, we compute the center of mass of the cell by iterating through all pieces of cytoplasm and computing the average $x$ and $y$ values. 
We then compute the distance from the spawn point using the distance formula with the cell's current center of mass and the spawn point (49, 49). Finally, we compute the total distance traveled by summing the distance between the current center of mass and the center of mass an iteration before. For each calculation, we have 2000 trials of each cell size.
 
\begin{figure}[H]
\begin{center}
\includegraphics[scale=0.13]{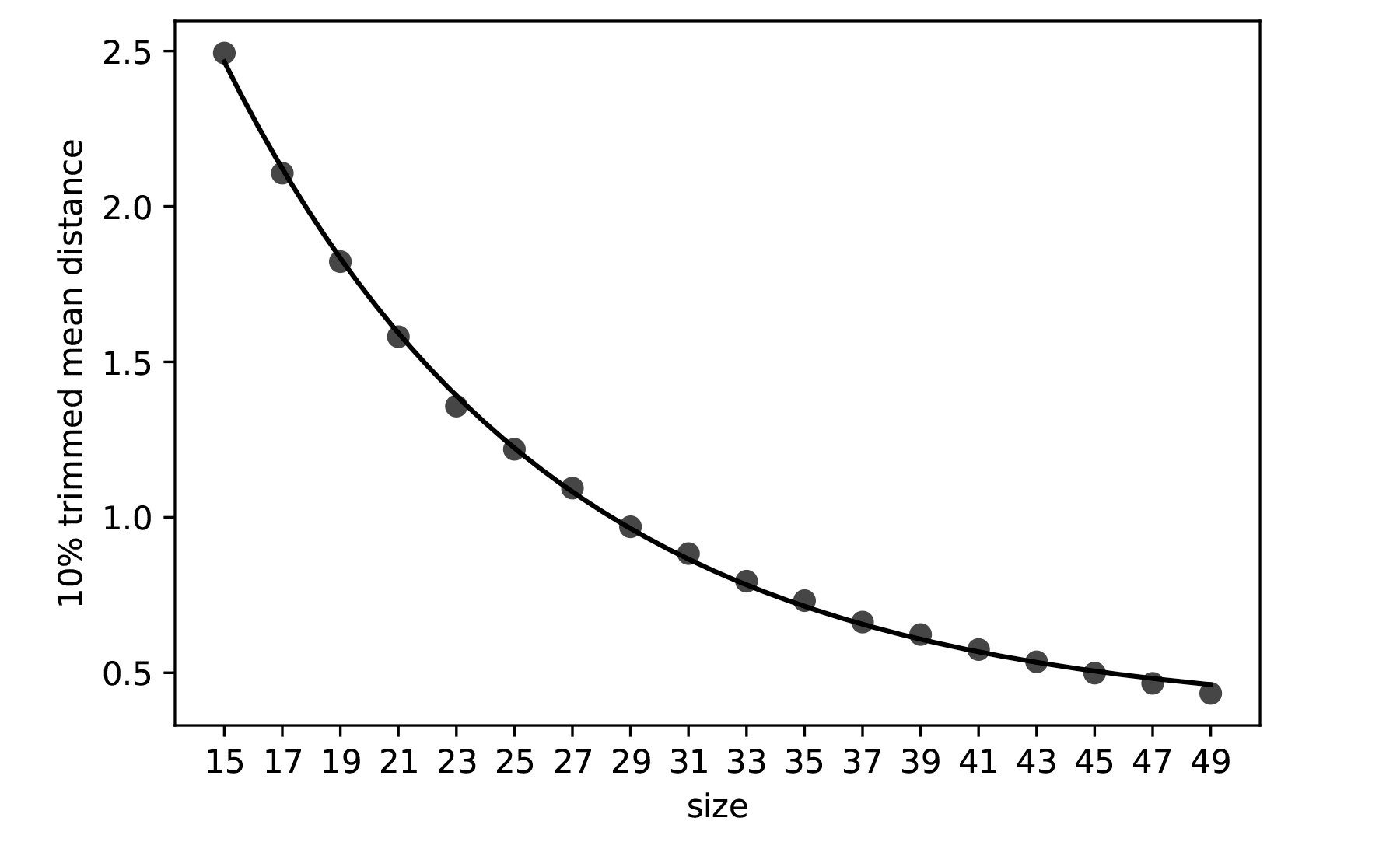}
\caption{We compute the ten percent trimmed mean of each cell size. We then fit an exponential curve to the data. %The resulting curve is \eqref{eq:av_dis_fit}
}
\label{fig:av_dis_exp_10_trim}
\end{center}
\end{figure}

  We shall begin by analyzing the average distance from the spawn point. To do this, we compute the ten percent trimmed mean of the average distance for each given cell size, and then fit an exponential curve to this data. The results are shown in Figure \ref{fig:av_dis_exp_10_trim}. The exponential curve (Eq. ~\eqref{eq:exponential}) $a =8.0305104, b=  0.91459648$, and $c = 0.36077371$ appears to fit the data points very well, suggesting that the relationship between the average distance and size resembles that of exponential decay.
  
\begin{figure}[H]
\begin{center}
\includegraphics[scale=0.52]{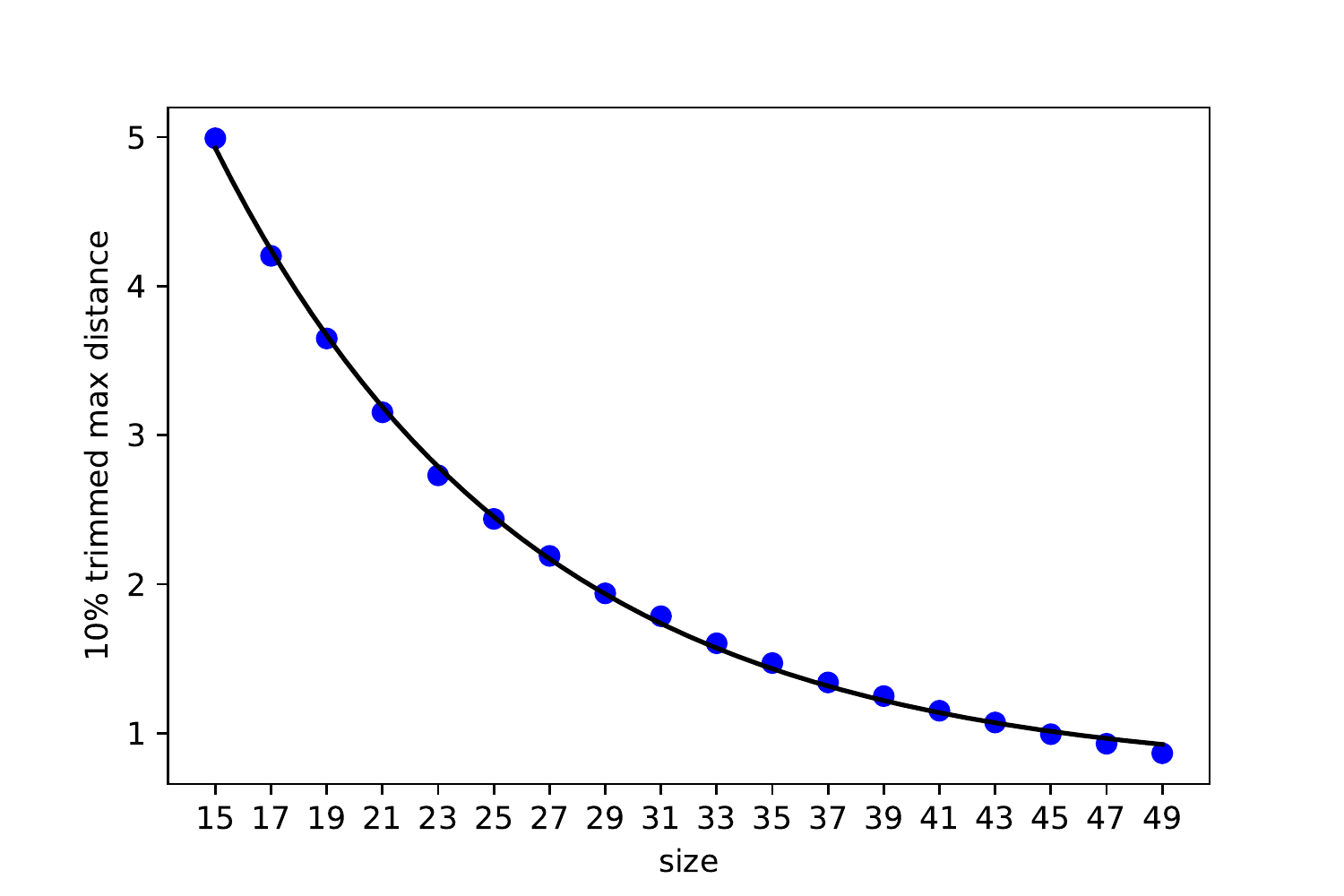}
\caption{We compute the ten percent trimmed mean of each cell size. We then fit an exponential curve to the data.}
\label{fig:ld_size_exp_10_trim}
\label{eq:ld_size_fit}
\end{center}
\end{figure}

\begin{figure}[H]
\begin{center}
\includegraphics[scale=0.6]{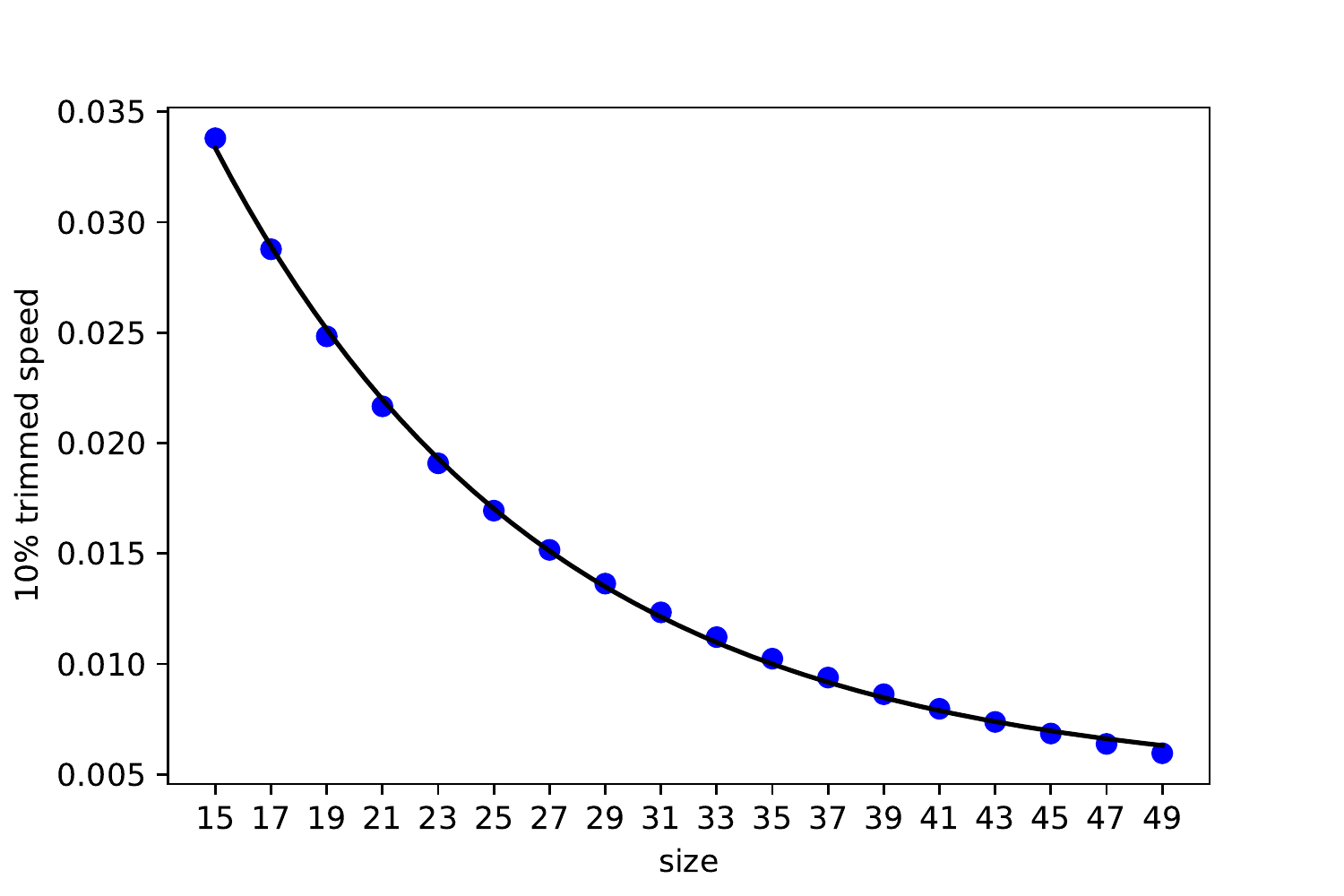}
\caption{We compute the 10 percent trimmed mean of the 1000 trials. We then attempt to fit the data to the exponential curve (Eq.~\eqref{eq:exponential}) with values $a = 0.10144046, b =  0.9192343$, and $c =   0.0046737$. }
\label{fig:speed_dis_exp_10_trim}
\label{eq:speed_dis_exp_10_trim}
\end{center}
\end{figure}

In what follows we shall consider the largest distance from spawn point:  this value represents the furthest point that a cell has traveled to.   We first compute the $10\%$ trimmed mean of the average distance for each given cell size. We then fit the exponential curve (Eq.~\eqref{eq:exponential} with values $a = 15.91026266, b = 0.91514874$, and $c =   0.71778398$, as  shown in Figure \ref{fig:ld_size_exp_10_trim}, allowing us to conclude that the relationship between the largest distance and area resembles that of exponential decay. 
%
%\begin{equation}
%
%\label{eq:ld_size_fit}
%\end{equation}
\newpage
Finally, we shall  analyze the speed of the fusion: at each iteration of our modified CELL model, we compute the distance of the center of mass from the previous center of mass. Averaging all the distances to find the average speed of the cell, we see that  the data fits an exponential curve, as shown in   Figure \ref{fig:speed_dis_exp_10_trim}. 
%\begin{figure}[H]
%\begin{center}
%\includegraphics[scale=0.5]{speed_dis_scatterplot}
%\includegraphics[scale=0.5]{speed_dis_boxplot}
%\caption{We graph the 1000 trials of each distance in a scatterplot and boxplot.}
%\label{fig:speed_dis}
%\end{center}
%\end{figure}
%\begin{equation}
%y = 0.11035326 \cdot 0.91923427^x + 0.0046737
%\label{eq:speed_dis_exp_10_trim}
%\end{equation}

\subsection{Cytoplasm mixing in cell fusion}
In this section we shall further consider how the sizes of cells can impact their fusion. To do so,  we spawn two cells of varying size and measure how well the cytoplasm of each cell mix together. We use Lacy's mixing index $M$,  described in \cite{plu13}, to measure the mixing of cytoplasm. Here, the variable $M$ ranges from 0 to 1,  where 0 is not mixed at all and 1 is completely mixed, and  is defined in Eq.~\eqref{eq:M} below: 
\begin{equation}
M = \frac{S_0^2 - S^2}{S_0^2-S_R^2}, 
\label{eq:M}
\end{equation}
where $S_0$ and $S_R$ are defined in \eqref{eq:S0} and \eqref{eq:SR} respectively. 
\begin{eqnarray}
S_0^2 &=& qp;
\label{eq:S0}
\\
S_R^2 &=& \frac{qp}{N},
\label{eq:SR}
\end{eqnarray}
for $q$  the proportion of the mixture that is of the first substance,  $p$  the proportion of the mixture that is of the second substance, and $N$  the number of particles in the samples. 

\begin{figure}[H]
\centering \includegraphics[scale=0.07]{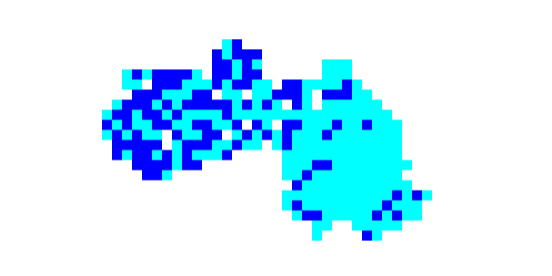}
\caption{Two cells of size 15 and 20 respectively. After 5000 iterations, mixing index of 0.726373742358510.}
\label{fig:mixing_example}
\end{figure}

The standard deviation over $n$ samples is shown in \eqref{eq:S}, where we define $x_i$ to be the proportion of the sample that is the first substance, and set $\bar{x}$ to equal $p$.
\begin{equation}
S^2 = \frac{1}{n}\sum_{i=1}^{i=n}(x_i - \bar{x})^2.
\label{eq:S}
\end{equation}
As a point of reference, Figure \ref{fig:mixing_example} has a mixing index of 0.726.

\begin{figure}[H]
\begin{center}
\includegraphics[scale=0.06]{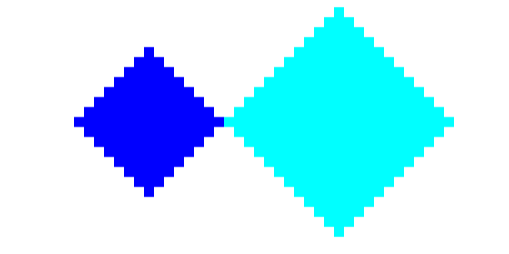}
\caption{Here we have two CELLs of size 15 (left) and 23 (right). The CELLs have just been spawned so that they are barely touching each other at the center of the image.}
\label{fig:sample_setup_mixing}
\end{center}
\end{figure}

Considering a starting  cell of size 15, we then spawn another cell of varying size from 15 to 49 (odd sizes only), and run 1000 trials for each of the sizes. Each trial consists of 10000 iterations of our modified CELL model. The cells are spawned so that they are initially barely touching as shown in Figure \ref{fig:sample_setup_mixing}. In each iteration, we compute the mixing index as discussed above. We sample the current state of the CELLs by taking all $3 \times 3$ squares in the grid. If more than half of the $3 \times 3$ square is not cytoplasm, we disregard the sample. For each of the trials, we compute the mean value of the mixing index as well as the largest mixing index.

\begin{figure}[H]
\begin{center}

\includegraphics[scale = 0.5]{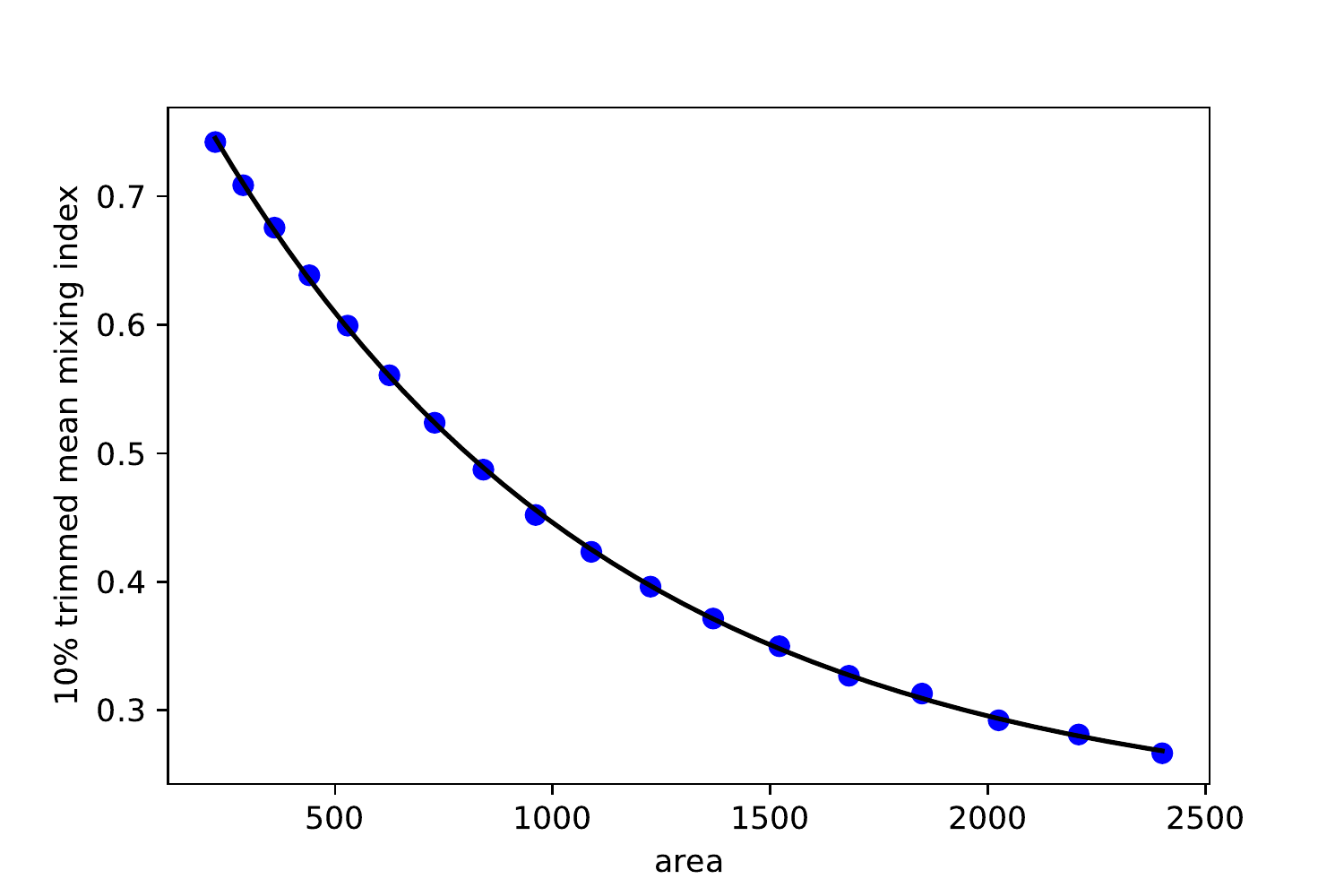}
\caption{We compute the 10 percent trimmed mean of the mean mixing index. We fit this data to an exponential function (Eq.~\eqref{eq:exponential}) pictured in blue with values $a = 0.67188919, b = 0.99891735$, and $c =  0.21859742$.}
\label{fig:mean_area_exp_10_trim}
\end{center}
\end{figure}

\begin{figure}[H]
\begin{center}
\includegraphics[scale = 0.5]{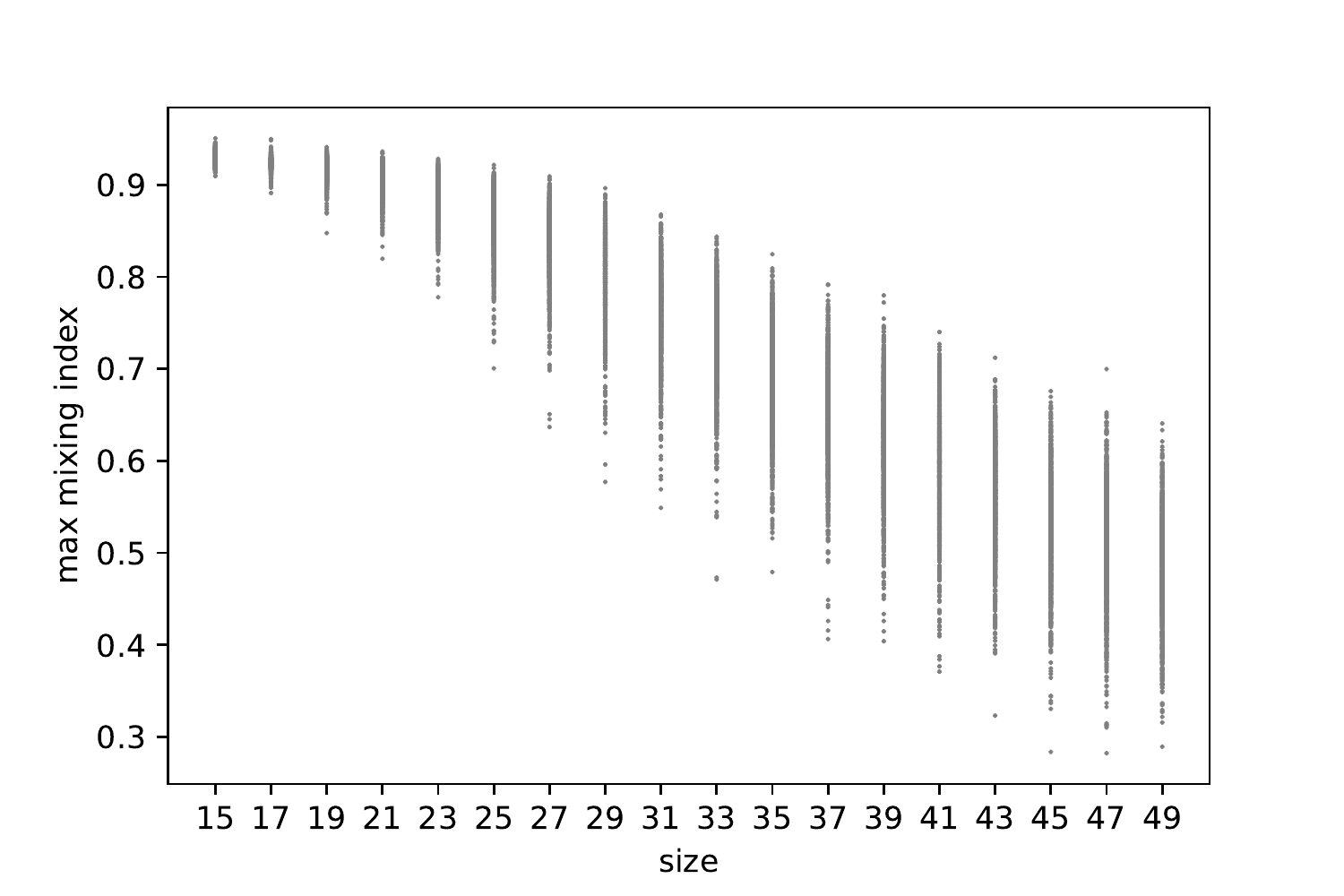}
\includegraphics[scale=0.5]{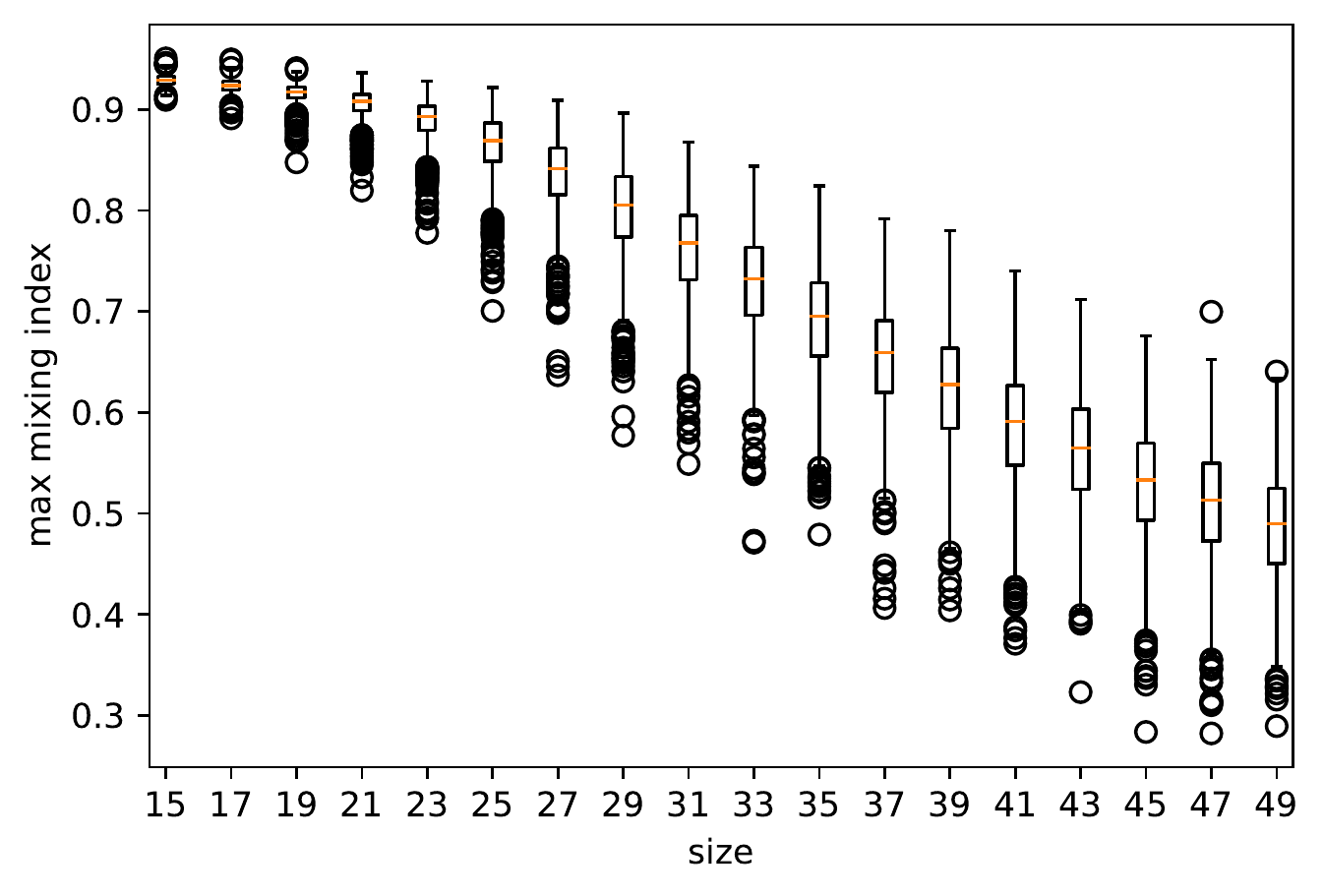}
\caption{We graph the data from our 1000 trials of each size ranging from 15 to 49 (odd only) using a scatterplot and boxplot.}
\label{fig:max_dis_all}
\end{center}
\end{figure}

\begin{figure}[H]
\begin{center}
\includegraphics[scale = 0.5]{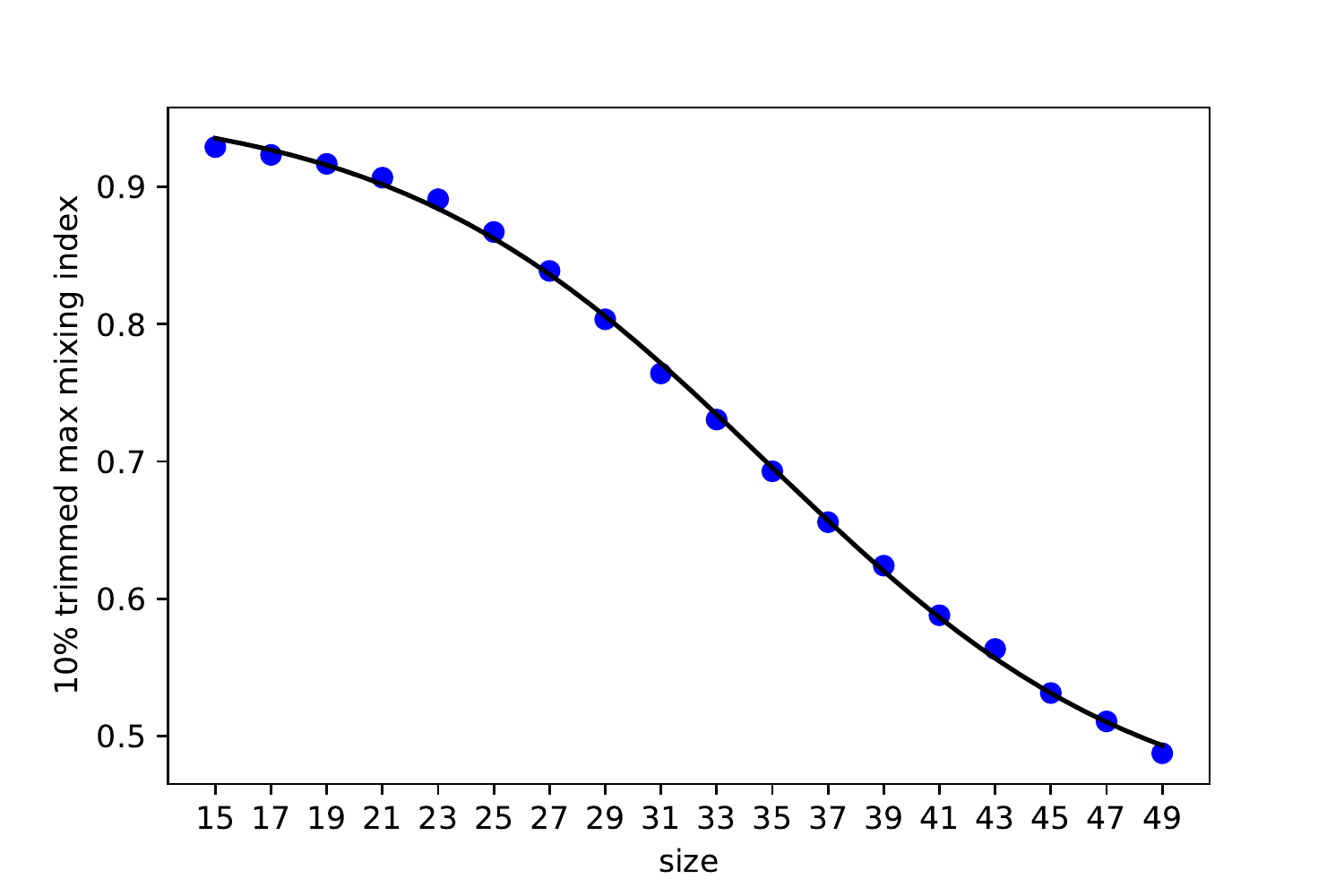}
\caption{We compute the 10 percent trimmed mean of the mean mixing index. We fit this data to a logistical function (Eq.~\eqref{eq:logistical}) pictured in blue. The parameters are $L = -0.52639764, k = 0.52639764, x_0 = 34.83393314$ and $b =0.96196691$.}
\label{fig:max_dis_log_10_trim}
\end{center}
\end{figure}

We shall first analyze the mean mixing index $M$ with respect to area, which is proportional to size, and which we have been analyzing in the previous sections, squared.  For each area size, we trim the lower and upper 10 percent of the dataset and graph the 10 percent trimmed mean in Figure \ref{fig:mean_area_exp_10_trim}. This data is clearly best fit by an exponential function.
We now analyze the maximum mixing index, or the highest amount of mixing achieved in the 10000 iterations.
We see in Figure \ref{fig:max_dis_all} that the max mixing index decreases as size increases and has more lower outlines than upper outlines.
Finally, we  compute the 10 percent trimmed mean of the max mixing index, and fit this data to a logistical function in Figure \ref{fig:max_dis_log_10_trim}.
\pagebreak
  \section{Concluding remarks}\label{last}
 In the present paper we have modified the original CELL model \cite{plu8} in order to be able to study  interactions between multiple CELLs of varying sizes and at different distances. Four different experiments were run. The first of these experiments measured the time it took for two cells to come into contact based on the distance between them. The second of these experiments tested which two cells out of three would merge first. The third of these experiments tested the mobility of cells in relation to cell size. The fourth of these experiments measured the mixing of cytoplasm of two cells in relation to cell size. 
 \bigskip 
 
\noindent {\bf Summary of results.} 
   Our observed results corroborate multiple expected \textit{Physarum} behaviors:  merging time increases with distance, cells merge with closer cells, cells are more mobile, and smaller cells fuse better. This further validates the CELL model as well as the modified CELL model with multiple CELLs of varying sizes and at varying distances. Furthermore, we have been able to identify the relationship between important characteristics such as speed vs distance (experimental decay) which would have been difficult to determine by experimentally growing \textit{Physarum}.
   \par
   From the first experiment, we were able to identify that the correlation between the time to contact and the distance between CELLs is linear. In the second experiment, we determined that CELLs most often merge with the CELLs closest to them and do so quicker than merging with other CELLs. In the third experiment, we determine that there is an exponential decay relationship between measures of mobility (speed, distance from start) and size. In the fourth experiment, we identify the relationship between the mean mixing index and area as exponential decay and the maximum mixing index and size as logistical decay.
   \bigskip
   
  \noindent {\bf Applications.} 
 We foresee many  applications of our work. In particular, our modified CELL model can be used in the following setups:
   \begin{itemize}
      \item[(a)] Fusion between multiple CELLs can also be used to model the rise of present day civilization from the early nomadic humans \cite{mig1}. Each tribe or group is modeled as a cell. Smaller groups or tribes are able to travel faster. As the size grows, they become less and less mobile until permanent settlement occurs. As the groups travel around, they meet other groups where they could merge. As with the results described in this paper, groups closer together have a closer connection than groups further apart.

   \item[(b)] Multiple CELLs can be used to model the spread of knowledge among groups of people. Following the ideas of \cite{bhansali2020trust}, certain groups would begin with certain knowledge, and when they encounter another group, they would combine and both share the knowledge possibly modified. In such way, one could potentially understand the spread and deformation of ideas and information. 

   \end{itemize}
\par   
    In addition, there is more biological work that can be done to verify these results and further validate the CELL model \cite{plu8} as well as our generalized CELL model. In particular, each of the experiments done with the CELL model can be done with physical \textit{Physarum Polycephalum}. In addition, multiple \textit{Physarum} organisms can share information and communicate prior to protoplasm fusion through the mucus that surrounds \textit{Physarum}. New models can be developed or existing models can be modified to represent this additional aspect of multiple \textit{Physarum} fusing.
    \par
    There is also potential to use multiple \textit{Physarum} organisms to solve problems a single \textit{Physarum} has already solved, such as maze solving and tree formation, and this is currently being studied in  \cite{soon} by the authors.
    \bigskip
    
\noindent{\bf Final comments.}
    Exploring the interaction between multiple \textit{Physarum Polycephalum} organisms is essential to discovering more about \textit{Physarum}'s unique ability as a unicellular organism to communicate with other cells and remember past events. More research into \textit{Physarum}'s unique learning, communication, and memory capabilities will allow us to discover more about the mysteries of human's and animal's brains and ability to learn, communicate, and remember.
    
    \bigskip

 \noindent {\bf Acknowledgments.} The  authors   are thankful to MIT
PRIMES-USA for the opportunity to conduct this research together, and in particular to Fidel A. Schaposnik for bringing   \textit{Physarum Polycephalum} to our attention after it started growing in his boiler. 
%her continued support and James Unwin for insightful comments on a draft of the manuscript.
  The work of Laura Schaposnik is partially supported through the NSF grant  CAREER DMS 1749013.   \\
 
 \noindent {\bf Affiliations.}\\
  (a)  Valley Christian High School, San Jose,   USA. \\
  (b)  University of Illinois, Chicago,  USA. \\

\bibliography{PRIMES2020}{}
\bibliographystyle{plain}

%\begin{thebibliography}{9}
% 
% 
%\bibitem{Henle}Henle
%\bibitem{Ross}Ross
%\bibitem{Kermack}Kermack
%\bibitem{markov}markov
%\bibitem{regression}regression
%\bibitem{charts}charts
%\bibitem{stochastic}stochastic
%\bibitem{differential}differential
%\bibitem{complex}complex
%\bibitem{spatial}spatial
%\bibitem{Galton}Galton
%
% \end{thebibliography}
% 

%\linenumbers


\end{document}